# Carrier-Specific Femtosecond XUV Transient Absorption of PbI$_2$ Reveals Ultrafast Nonradiative Recombination

*Ming-Fu Lin[1,3,4], Max Verkamp[1], Joshua Leveillee[2], Elizabeth Ryland[1], Kristin Benke[1], Kaili Zhang[1], Clemens Weninger[3,4], Xiaozhe Shen[5], Renkai Li[5], David Fritz[3], Uwe Bergmann[4], Xijie Wang[5], André Schleife[2], and Josh Vura-Weis[1*]*

[1] Department of Chemistry, University of Illinois at Urbana-Champaign, Urbana, IL 61801

[2] Department of Materials Science and Engineering, University of Illinois at Urbana-Champaign, Urbana, IL 61801

[3] Linac Coherent Light Source, SLAC National Accelerator Laboratory, Menlo Park, CA 94025

[4] Stanford PULSE Institute, SLAC National Accelerator Laboratory, Menlo Park, CA 94025

[5] SLAC National Accelerator Laboratory, Menlo Park, CA 94025

**Corresponding Author**

*vuraweis@illinois.edu

## ABSTRACT

Femtosecond carrier recombination in PbI$_2$ is measured using tabletop high-harmonic extreme ultraviolet (XUV) transient absorption spectroscopy and ultrafast electron diffraction. XUV absorption from 45 eV to 62 eV measures transitions from the iodine 4*d* core level to the conduction band density of states. Photoexcitation at 400 nm creates separate and distinct transient absorption signals for holes and electrons, separated in energy by the 2.4 eV band gap of the semiconductor. The shape of the conduction band and therefore the XUV absorption spectrum is temperature dependent, and nonradiative recombination converts the initial electronic excitation to thermal excitation within picoseconds. Ultrafast electron diffraction (UED) is used to measure the lattice temperature and confirm the recombination mechanism. The XUV and UED results support a 2$^{nd}$-order recombination model with a rate constant of $2.5 \times 10^{-9}$ cm$^3$/s.

## TOC GRAPHICS

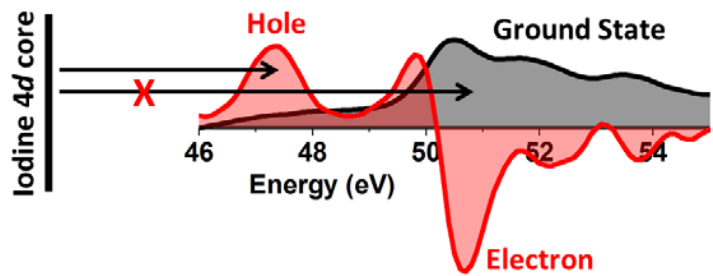

**1. INTRODUCTION**

Ultrafast dynamics of light-induced carriers are central to the function of photovoltaic and photocatalytic semiconductors.[1] In this work, we use femtosecond XUV transient absorption to measure carrier relaxation dynamics in $PbI_2$, and show that the transient spectrum of wide-bandgap semiconductors produces distinct signals for photoinduced electrons and holes. The conduction band shape and therefore the XUV spectrum is sensitive to temperature, and nonradiative recombination to lattice heat is observed on a ~4 ps timescale. Ultrafast electron diffraction is used to quantify the final sample temperature and identify the primarily nonradiative recombination pathway.

A variety of spectroscopic methods have historically been used to measure the rate and mechanism of carrier cooling, recombination, and transport. Transient UV/Visible absorption and photoluminescence spectroscopy probe photoinduced changes in the band structure such as bandgap renormalization and band broadening, as well as the population and energy distribution of carriers in each band.[2–7] In those experiments the electron and hole dynamics are measured together as a single observable, namely the time-resolved absorption or luminescence spectrum, and it can be difficult to disentangle the electron vs. hole dynamics. In contrast to those band-to-band probes, transient infrared spectroscopy of semiconductors probes intraband transitions, and in some cases can distinguish between electron and hole relaxation.[8–10] Terahertz spectroscopy measures the carrier density and mobility, and is especially useful for identifying free carriers.[11]

Finally, time-resolved two-photon photoemission selectively probes electron dynamics, independent of hole relaxation.[12]

Extreme ultraviolet (XUV) transient absorption spectroscopy is emerging as a powerful tool for measuring femtosecond to attosecond phenomena in solid-state systems. XUV photons in the 30 eV to 100 eV spectral range induce core-to-valence transitions and therefore provide an element-specific probe of electron dynamics. XUV transient absorption of semiconductors[13] has been used to measure charge transfer dynamics in transition metal oxides[14–16] and attosecond bandgap renormalization upon strong-field perturbation in Si.[17] This technique was recently used to measure the few-femtosecond electron and hole relaxation in Ge[18] and Ge/Si alloys,[19] revealing carrier- and valley-specific relaxation. The small bandgap and significant core-hole spin-orbit coupling of those semiconductors led to overlapping hole and electron signals, which were separated using a sophisticated iterative algorithm. In the present work, we show that this complexity is largely removed for wide band-gap semiconductors, providing a simple and intuitive picture of the hole and electron dynamics. Furthermore, access to the I 4d core level provides a foundation for future studies of lead halide perovskites such as $CH_3NH_3PbI_3$. $PbI_2$ itself is a synthetic precursor for these materials, and residual $PbI_2$ can have either a positive or negative effect on device performance, depending on the concentration and morphology.[20–25] The relaxation rate and mechanism of photoinduced carriers in $PbI_2$ are therefore important factors in the performance of such devices. Previous UV/visible transient absorption studies of $PbI_2$ films have observed fast (~6 ps) and slow (>30 ps) decay components, but reached differing conclusions regarding the relaxation mechanism. In one work the fast component was assigned as electron trapping and the slow component as carrier recombination.[26] In the other, the fast component was assigned as carrier recombination with the long component ascribed to heat dissipation.[27]

## 2. METHODS

Visible-pump/XUV-probe transient absorption spectroscopy was performed using an instrument diagrammed in Figure 1A.[28] A 1 kHz, 4 mJ, 35 fs, 800 nm pulse was focused into a semi-infinite gas cell[29] filled with ~70 torr of argon gas, with a peak intensity at the focal point of $6\times10^{14}$ W/cm$^2$. The strong oscillating electric field ionizes the argon atoms, accelerates the electrons, then drives them back into the nuclei, producing a broadband (45 eV to 62 eV) XUV pulse via the process of high-harmonic generation.[30] The XUV beam was focused by a toroidal mirror to a ~100 µm FWHM spot at the sample in transmission geometry, then dispersed by a homemade spectrometer onto a CCD area detector. The spectrometer resolution was measured as 0.4 eV FWHM using the atomic absorption lines of Xe and Xe$^+$, and the spectra were binned in 0.1 eV increments. The spectrum of the XUV continuum, as detected at the CCD after transmission through the PbI$_2$ sample, is shown in Figure 1B. Peaks in the spectrum correspond to odd harmonics of the 800 nm driving laser field, but a broad continuum beneath these peaks allows the full 45 eV to 62 eV energy range to be used for spectroscopy. Transient absorption spectra were collected by photoexciting the PbI$_2$ film at 3.1 eV (400 nm) with a peak fluence of ~2.9 mJ/cm$^2$ and measuring the differential absorption signal $\Delta A = log_{10}(I_{off}/I_{on})$, where I$_{off}$ and I$_{on}$ are the XUV spectra with pump off and pump on, respectively. This pump fluence produces a charge carrier density of ~$1.5\times10^{20}$ cm$^{-3}$, corresponding to the electron-hole plasma regime. The sample film was raster-scanned to avoid pump-induced damage and cooled by flowing N$_2$ gas to avoid accumulative sample heating.

## 3. RESULTS AND DISCUSSION

### 3.1: XUV Transient Absorption Spectra

The ground-state XUV absorption spectrum of a ~70 nm thick $PbI_2$ polycrystalline film is shown in Figure 2A (black curve), and closely matches the spectrum of single-crystal $PbI_2$ collected using a synchrotron radiation light source.[31] Major peaks are observed at 50.5 eV, 51.6 eV, and 53.5 eV, respectively. To first approximation, this spectrum maps the contribution of iodine orbitals to the conduction band (CB) density of states, convolved with the 1.7 eV spin-orbit splitting of the iodine $4d$ core hole ($4d_{5/2}$, $4d_{3/2}$) spin-orbit states.[32] An abbreviated band structure of $PbI_2$ calculated using density functional theory with spin-orbit coupling is shown in Figure 2B. The full band structure,[33] along with computational details, is given in the Supporting Information. The valence band maximum is composed of Pb 6s and I 5p orbitals, with 83% I character, while the conduction band minimum is formed from Pb 6p and I 5p with 38% I character. XUV transitions from the I 4d core to the first three conduction bands are shown as vertical arrows in Figure 2B, and optical transition dipole matrix elements are summed across all $k$ points. The calculated transitions were broadened with a 0.4 eV FWHM Gaussian and a 0.5 eV FWHM Lorentzian to account for the spectrometer resolution and core-hole lifetime broadening, respectively. This lifetime broadening is estimated from the width of the iodine $4d$ photoelectron spectrum, as discussed in the Supporting Information. Finally, the core-level binding energy is underestimated at this level of theory, so the simulated spectrum is shifted manually by +4.7 eV to match the experimental result.

The simulated spectrum shown in Figure 2A is a qualitative match to the experimental spectrum. The overall shape and peak-to-peak spacing of the experiment is well reproduced, although the height of the 50.5 eV peak is underestimated by the calculation, possibly due to the neglect of excitonic effects. Also, while the degeneracy of the core hole total angular momentum J states would predict a 3 to 2 ratio between the $4d_{5/2} \rightarrow$ CB and $4d_{3/2} \rightarrow$ CB transitions (solid vs dashed lines in Figure 2), spin-orbit coupling in the conduction bands leads to spin- and band-dependent transition dipole moments. Ratios of 3.3 to 1, 2.1 to 1, and 1.2 to 1 between the $4d_{5/2} \rightarrow$ CB and $4d_{3/2} \rightarrow$ CB transitions are predicted for the three conduction bands, respectively. A detailed treatment of the interaction between core and valence spin-orbit coupling as well as excitonic effects is beyond the scope of this paper, but will be explored in depth in a future publication.

Transient XUV absorption spectroscopy after excitation of the direct band-to-band transition at 400 nm produces the time-resolved difference spectra shown in Figure 3A. At early times (<1 ps), the transient spectrum is characterized by a positive feature at 47.4 eV and a derivative-shaped feature with positive and negative peaks at 49.9 eV and 50.7 eV, respectively. Minor negative peaks are observed at 52.1 eV and 53.8 eV. Over the next few ps, the 47.4 eV peak disappears, the positive peak at 49.9 eV doubles in intensity, and a broad positive feature emerges from 55 eV to 62 eV. The initial and final spectra are shown in Figure 3B, averaged from 0.05 ps to 0.25 ps and from 40 ps to 100 ps, respectively.

Kinetic traces at 47.4 eV and 57.0 eV are shown in Figure 4. Singular Value Deposition and global fitting of the entire data set is performed using a sequential A$\rightarrow$B model and either first- or second-order order kinetics ($-\frac{d[A]}{dt} = \frac{d[B]}{dt} = k_n[A]^n$, where $n$ is 1 or 2). As shown in the figure, both kinetic models fit the data well within the noise of the experiment. The species-associated spectra obtained from the two global fits are nearly identical to each other and to the initial and final spectra in Figure 3B, as shown in Figure S8. The first-order fit gives a 4.2 ± 0.2 ps exponential decay time constant ($1/k_1$). The second-order fit gives a rate constant of $k_2 = 2.5\times10^{-9}$ cm$^3$/s, assuming an initial carrier concentration of $1.5\times10^{20}$ cm$^{-3}$ as calculated from the pump fluence. Both fits give a 95 fs FWHM instrument response function. The physical implications of each model will be discussed in Section 4

**3.2 Interpretation of initial transient spectrum: distinct signals for holes and electrons**

The initial transient spectrum (0.05 ps to 0.25 ps) shown in Figure 3B is interpreted using the model shown in Figure 5A. The positive feature at ~47.4 eV arises from valence band holes. Photoexcitation with the 3.1 eV pump pulse creates a Fermi distribution of holes, opening a channel for XUV absorption from the I 4$d$ core to these newly-unoccupied states. This new channel is observed as a positive transient absorption signal that is redshifted by approximately the 2.4 eV band gap[34] from the conduction band minimum (Figure 5B, yellow area). The integrated area of this signal is proportional to the hole population, rising with the pump pulse then decaying in a few ps as the holes recombine with electrons.

The derivative-shaped feature at ~50 eV is caused by conduction band electrons via a combination of bandgap renormalization (BGR) and band filling (BF).[2,35,36] As shown in Figure 5A, BGR shifts the conduction band to lower energy, which would cause the derivative-shaped transient in Figure 5B (red and blue areas). However, BF due to the photogenerated electron population blocks absorption into the bottom of the shifted band, reducing the intensity of the positive signal (red area). The magnitude of this positive signal depends on the relative magnitudes of the BGR vs BF. The simulation shown in Figure 5B results from a 30 meV shift and 1.7% filling of the first conduction band. These numbers are reasonable given the $1.5\times10^{20}$

cm$^{-3}$ photoinduced carrier density, as discussed in the Supporting Information. Simulations of the transient spectrum considering only BGR or only BF are shown in Figures S13 and S14.

Finally, spin-orbit coupling of the I 4$d$ core-hole is incorporated in the simulation. The transient signal from I 4$d_{3/2}$ to the valence and conduction bands will appear 1.7 eV higher in energy than the 4$d_{5/2}$ → VB/CB signals. As was shown in Figure 2B, the intensity ratio between the 4$d_{5/2}$ and 4$d_{3/2}$ signals is band-dependent and differs from the 3 to 2 value expected from the core-hole degeneracy due to spin-orbit dependent transition dipole moments. The DFT calculation described above predicts a 4$d_{5/2}$ to 4$d_{3/2}$ ratio of 3.3 to 1 for the transition to the lowest conduction band, and 5.0 to 1 for the transition to the highest valence band. The VB and CB features in Figure 5B are shifted and scaled accordingly to give the 4$d_{3/2}$ contribution to the spectrum, giving the total spectrum shown in Figure 5C. This simulation is a good match for the experimental initial transient spectrum, as shown in Figure 5D. Note that in Figure 5A-C, the conduction band minimum is set as the zero of energy, while in Figure 5D the conduction band minimum is set to the rising edge of the ground-state spectrum 49.8 eV.

While this simple model accounts for most of the observed features in the initial transient spectrum, two limitations should be noted. First, the effect of bandgap renormalization of higher conduction bands (CB2 and CB3 in Figure 2) is not included, nor is photoinduced broadening of the three conduction bands. A redshift of the upper bands would cause additional derivative-shape signal in the transient spectrum from 51 eV to 55 eV, which are in fact observed in the experiment. Accurate simulation of the transient spectrum in this region is difficult because of the uncertain magnitude of the shift and/or broadening of the upper conduction bands, as well as overlap with I 4$d_{3/2}$→CB transitions. We have therefore not attempted to model these effects, which would risk overfitting the spectrum. Second, a shift in the I 4$d$ core is likely upon photoexcitation, but is expected to be small due to the delocalization of the carriers at these short timescales.[37] Photoexcitation of $PbI_2$ causes a net shift in electron density from I to Pb,[33] which would lower the 4*d* core and blueshift the core→VB transition. This shift partially counteracts the observed effect of bandgap renormalization, so the 30 meV conduction band shift in Figure 5 should be interpreted as the sum of the bandgap renormalization and the 4*d* shift.

### 3.3 Interpretation of final transient spectrum: lattice heat

The final transient spectrum (40 ps to 100 ps) is characterized by a derivative-shaped transient spectrum at the conduction band edge and a broad positive absorption feature from 55 eV to 62 eV. This spectrum is the effect of increased lattice disorder caused by carrier recombination (i.e. heat). Carrier relaxation heats the lattice via phonon emission, shrinking the band gap and modifying the conduction band density of states.[1] Two experiments were performed to support this assignment. First, the absorption spectrum of a $PbI_2$ sample heated to 120±10 °C was compared to that of a 20 °C sample, resulting in a difference spectrum that strongly resembles the late-time spectrum as shown in Figure 6. This similarity suggests that the final transient spectrum is caused by thermal effects, as opposed to a long-lived electronic state.

Second, ultrafast electron diffraction (UED) with ~250 fs time resolution was performed to confirm the timescale and magnitude of sample heating.[38–40] This technique measures dynamic structural changes as a result of the lattice temperature jump via the Debye-Waller response.[38–40] Figure 7A shows the diffraction pattern of a 180 nm thick $PbI_2$ film.[41] The change in intensity of the (300) and (410) peaks after 3.1 eV photoexcitation with an initial carrier density of $0.33 \times 10^{20}$ cm$^{-3}$ is shown in Figure 7B,C. A lower carrier density was used here compared to the XUV experiment to avoid long-lived sample heating, as the nitrogen gas flow used to cool the sample between laser shots in the XUV experiment was unavailable in the UED experiment. The 180 Hz UED repetition rate (vs 1 kHz for XUV) also helps to cool the sample between shots. The time-dependent Bragg peak intensity drop can be fit to either first-order or second-order kinetic models. For the first-order model, time constants of 15.1±1.2 ps and 11.2±0.9 ps are found for the (300) and (410) peaks, respectively. For the second-order model, rate constants of $2.4\times10^{-9}$ cm$^3$/s and $3.4\times10^{-9}$ cm$^3$/s are obtained for the (300) and (410) peaks. As shown in Figures 7B and 7C, the second-order fit is considerably better than the first-order fit, especially in the 10 ps to 100 ps time window. The second-order rate constants are in good agreement with the $2.5\times10^{-9}$ cm$^3$/s rate constant obtained from the transient XUV data, implying that the decay of the early-time XUV signal and the appearance of sample lattice disorder arise from the same underlying mechanism.

Finally, the Debye-Waller response at long delay times is analyzed to quantify the heat deposited in the lattice. Figure 8 shows this response as a function of $Q^2$, along with the predicted response for two limiting cases. In the nonradiative recombination limit, each pump photon would deposit 3.1 eV of thermal energy into the lattice. In the radiative limit, carrier relaxation to the band edge would deposit 0.7 eV of thermal energy, with the final 2.4 eV band-gap energy released as an emitted photon. The close match between the experimental result and the nonradiative limit prediction shows that the majority recombination pathway is in fact nonradiative. Given this mechanism and the known pump fluence, the estimated temperature rise of the transient XUV experiment is 133K, consistent with the good match between the 120 °C spectrum in Figure 6 and the final transient spectrum. This calculation is described in detail in the Supporting Information.

## 4. Mechanism of carrier recombination

As discussed above, both the XUV and UED results show that carrier recombination is primarily nonradiative, with kinetics that can be fit to both first-order and second-order models with regard to excitation density. The $10^{20}$ cm$^{-3}$ photoinduced carrier density is much higher than the $10^{17}$ cm$^{-3}$ trap density in PbI$_2$ and other polycrystalline films.[42,43] Saturation of trap states therefore makes a first-order Shockley-Reed-Hall[44] trap-assisted recombination model physically unlikely. Second-order Auger recombination has been observed in amorphous silicon[45] and layered materials such as MoS$_2$[46] when there is a high density of overlapping electron-hole pairs ($\gtrapprox 10^{19}$ cm$^{-3}$). In this excitation regime, one exciton recombines and transfers its energy to a nearby exciton, which then dissociates into free carriers and dissipates the excess energy via phonon emission.[45,47] The average distance between excitons in the transient XUV and UED experiments is 23 Å to 27 Å. This value is close to the PbI$_2$ Bohr radius of 19 Å,[48] making the 2$^{nd}$-order Auger process the likely recombination mechanism. The $2.5\times10^{-9}$ cm$^3$s$^{-1}$ 2$^{nd}$-order rate constant measured with XUV transient absorption matches the $2.4\times10^{-9}$ cm$^3$s$^{-1}$ biexciton formation rate measured via power-dependent luminescence spectroscopy,[49] further supporting this assignment.

## 5. Conclusion

In summary, XUV transient absorption and ultrafast electron diffraction were used to measure photoinduced carrier dynamics in PbI$_2$ and identify a 2$^{nd}$-order nonradiative recombination mechanism. Core-level spectroscopy offers a unique handle for measuring the photophysics of semiconductors with multiple heavy atoms due to its element specificity. Hard x-ray absorption was recently used to measure the picosecond change in electron density and structural relaxation of each element in CsPb(Cl$_x$Br$_{1-x}$)$_3$ perovskite nanocrystals,[37] and free-electron lasers will enable such studies to be performed on femtosecond timescales. As shown in the current work, tabletop XUV transient absorption spectroscopy retains the element specificity of hard x-ray absorption while providing a straightforward mapping of the unoccupied valence and conduction band density of states. The presence of distinct signals for holes and electrons in the XUV region is especially powerful, as the dynamics of these carriers are often convolved in transient UV/visible spectroscopy. The combination of element- and carrier-specific spectroscopy demonstrated here opens the possibility of tracking electron and hole relaxation across semiconductor heterojunctions on femtosecond timescales. Finally, the Iodine $N_{4,5}$ edge at ~50 eV provides a convenient handle for measuring the photophysics of I-containing semiconductors such as methylammonium lead halide perovskites, and will allow carrier-specific studies of this important class of photovoltaic devices.

## ASSOCIATED CONTENT

**Support information**. Sample preparation, experimental details for XUV and UED measurements, Debye-Waller response analysis, computational details and full band structure diagram.

## AUTHOR INFORMATION

The authors declare no competing financial interests.

## ACKNOWLEDGMENT

This material is based upon work supported by the Air Force Office of Scientific Research under AFOSR Award No. FA9550-14-1-0314 to J.V.W. Density functional theory results are based upon work supported by the National Science Foundation under Grant No. CBET-1437230. This research is part of the Blue Waters sustained-petascale computing project, which is supported by the National Science Foundation (awards OCI-0725070 and ACI-1238993) and the state of Illinois. Blue Waters is a joint effort of the University of Illinois at Urbana-Champaign and its National Center for Supercomputing Applications. The authors would like to thank Prof. Aaron Lindeberg, Dr. Xiaoxi Wu and Dr. Ehern Mannebach for the discussion of data analysis in electron diffraction and interpretations. AFM, SEM and XRD measurements were carried out in part in the Frederick Seitz Materials Research Laboratory Central Research Facilities, University of Illinois. The UED work is performed at SLAC MeV-UED, which is supported in part by the DOE BES Scientific User Facilities Division and SLAC UED/UEM program development fund under the contract No. DE-AC02-05CH11231. This work was supported as part of the Computational Materials Sciences Program funded by the U.S. Department of Energy, Office of Science, Basic Energy Sciences, Materials Sciences and Engineering Division under Award Number DE-SC00014607

**REFERENCES**

(1)     Shah, J. *Ultrafast Spectroscopy of Semiconductors and Semiconductor Nanostructures*; Springer Series in Solid-State Sciences; Springer Berlin Heidelberg: Berlin, Heidelberg, 1999; Vol. 115.

(2)     Yang, Y.; Ostrowski, D. P.; France, R. M.; Zhu, K.; van de Lagemaat, J.; Luther, J. M.; Beard, M. C. Observation of a Hot-Phonon Bottleneck in Lead-Iodide Perovskites. *Nat. Photonics* **2015**, *10* (1), 53–59.

(3)     Zhu, H.; Miyata, K.; Fu, Y.; Wang, J.; Joshi, P. P.; Niesner, D.; Williams, K. W.; Jin, S.; Zhu, X.-Y. Screening in Crystalline Liquids Protects Energetic Carriers in Hybrid

Perovskites. *Science* **2016**, *353* (6306), 1409–1413.

(4) Alivisatos, A. P.; Harris, A. L.; Levinos, N. J.; Steigerwald, M. L.; Brus, L. E. Electronic States of Semiconductor Clusters: Homogeneous and Inhomogeneous Broadening of the Optical Spectrum. *J. Chem. Phys.* **1988**, *89* (7), 4001–4011.

(5) Elsaesser, T.; Shah, J.; Rota, L.; Lugli, P. Initial Thermalization of Photoexcited Carriers in GaAs Studied by Femtosecond Luminescence Spectroscopy. *Phys. Rev. Lett.* **1991**, *66* (13), 1757–1760.

(6) Shank, C. V; Fork, R. L.; Leheny, R. F.; Shah, J. Dynamics of Photoexcited GaAs Band-Edge Absorption with Subpicosecond Resolution. *Phys. Rev. Lett.* **1979**, *42* (2), 112–115.

(7) Ando, M.; Yazaki, M.; Katayama, I.; Ichida, H.; Wakaiki, S.; Kanematsu, Y.; Takeda, J. Photoluminescence Dynamics due to Biexcitons and Exciton-Exciton Scattering in the Layered-Type Semiconductor $PbI_2$. *Phys. Rev. B* **2012**, *86* (15), 155206.

(8) Narra, S.; Chung, C.-C.; Diau, E. W.-G.; Shigeto, S. Simultaneous Observation of an Intraband Transition and Distinct Transient Species in the Infrared Region for Perovskite Solar Cells. *J. Phys. Chem. Lett.* **2016**, *7* (13), 2450–2455.

(9) Burda, C.; Link, S.; Mohamed, M.; El-Sayed, M. The Relaxation Pathways of CdSe Nanoparticles Monitored with Femtosecond Time-Resolution from the Visible to the IR: Assignment of the Transient Features by Carrier Quenching. *J. Phys. Chem. B* **2001**, *105* (49), 12286–12292.

(10) Klimov, V. I.; Schwarz, C. J.; McBranch, D. W.; Leatherdale, C. A.; Bawendi, M. G. Ultrafast Dynamics of Inter- and Intraband Transitions in Semiconductor Nanocrystals:

Implications for Quantum-Dot Lasers. *Phys. Rev. B* **1999**, *60* (4), R2177–R2180.

(11) Ulbricht, R.; Hendry, E.; Shan, J.; Heinz, T. F.; Bonn, M. Carrier Dynamics in Semiconductors Studied with Time-Resolved Terahertz Spectroscopy. *Rev. Mod. Phys.* **2011**, *83* (2), 543–586.

(12) Niesner, D.; Zhu, H.; Miyata, K.; Joshi, P. P.; Evans, T. J. S.; Kudisch, B. J.; Trinh, M. T.; Marks, M.; Zhu, X.-Y. Persistent Energetic Electrons in Methylammonium Lead Iodide Perovskite Thin Films. *J. Am. Chem. Soc.* **2016**, *138* (48), 15717–15726.

(13) Borja, L. J.; Zürch, M.; Pemmaraju, C. D.; Schultze, M.; Ramasesha, K.; Gandman, A.; Prell, J. S.; Prendergast, D.; Neumark, D. M.; Leone, S. R. Extreme Ultraviolet Transient Absorption of Solids from Femtosecond to Attosecond Timescales. *J. Opt. Soc. Am. B* **2016**, *33* (7).

(14) Vura-Weis, J.; Jiang, C.-M.; Liu, C.; Gao, H.; Lucas, J. M.; de Groot, F. M. F.; Yang, P.; Alivisatos, A. P.; Leone, S. R. Femtosecond $M_{2,3}$-Edge Spectroscopy of Transition-Metal Oxides: Photoinduced Oxidation State Change in α-$Fe_2O_3$. *J. Phys. Chem. Lett.* **2013**, *4* (21), 3667–3671.

(15) Jiang, C.-M.; Baker, L. R.; Lucas, J. M.; Vura-Weis, J.; Alivisatos, A. P.; Leone, S. R. Characterization of Photo-Induced Charge Transfer and Hot Carrier Relaxation Pathways in Spinel Cobalt Oxide ($Co_3O_4$). *J. Phys. Chem. C* **2014**, *118* (39), 22774–22784.

(16) Carneiro, L. M.; Cushing, S. K.; Liu, C.; Su, Y.; Yang, P.; Alivisatos, A. P.; Leone, S. R. Excitation-Wavelength-Dependent Small Polaron Trapping of Photoexcited Carriers in α-$Fe_2O_3$. *Nat Mater* **2017**, *16* (8), 819–825.

(17) Schultze, M.; Ramasesha, K.; Pemmaraju, C. D.; Sato, S. A.; Whitmore, D.; Gandman, A.; Prell, J. S.; Borja, L. J.; Prendergast, D.; Yabana, K.; Neumark, D. M.; Leone, S. R. Attosecond Band-Gap Dynamics in Silicon. *Science* **2014**, *346* (6215), 1348–1352.

(18) Zürch, M.; Chang, H.-T.; Borja, L. J.; Kraus, P. M.; Cushing, S. K.; Gandman, A.; Kaplan, C. J.; Oh, M. H.; Prell, J. S.; Prendergast, D.; Pemmaraju, C. D.; Neumark, D. M.; Leone, S. R. Direct and Simultaneous Observation of Ultrafast Electron and Hole Dynamics in Germanium. **2017**, *8*, 15734.

(19) Zürch, M.; Chang, H.-T.; Kraus, P. M.; Cushing, S. K.; Borja, L. J.; Gandman, A.; Kaplan, C. J.; Oh, M. H.; Prell, J. S.; Prendergast, D.; Pemmaraju, C. D.; Neumark, D. M.; Leone, S. R. Ultrafast Carrier Thermalization and Trapping in Silicon-Germanium Alloy Probed by Extreme Ultraviolet Transient Absorption Spectroscopy. *Struct. Dyn.* **2017**, *4* (4), 44029.

(20) Chen, Q.; Zhou, H.; Song, T.-B.; Luo, S.; Hong, Z.; Duan, H.-S.; Dou, L.; Liu, Y.; Yang, Y. Controllable Self-Induced Passivation of Hybrid Lead Iodide Perovskites toward High Performance Solar Cells. *Nano Lett.* **2014**, *14* (7), 4158–4163.

(21) Wang, H.-Y.; Hao, M.-Y.; Han, J.; Yu, M.; Qin, Y.; Zhang, P.; Guo, Z.-X.; Ai, X.-C.; Zhang, J.-P. Adverse Effects of Excess Residual $PbI_2$ on Photovoltaic Performance, Charge Separation, and Trap-State Properties in Mesoporous Structured Perovskite Solar Cells. *Chem. - A Eur. J.* **2017**, *23* (16), 3986–3992.

(22) Liu, F.; Dong, Q.; Wong, M. K.; Djurišić, A. B.; Ng, A.; Ren, Z.; Shen, Q.; Surya, C.; Chan, W. K.; Wang, J.; Ng, A. M. C.; Liao, C.; Li, H.; Shih, K.; Wei, C.; Su, H.; Dai, J. Is Excess $PbI_2$ Beneficial for Perovskite Solar Cell Performance? *Adv. Energy Mater.* **2016**,

*6* (7), 1502206.

(23) Cao, D. H.; Stoumpos, C. C.; Malliakas, C. D.; Katz, M. J.; Farha, O. K.; Hupp, J. T.; Kanatzidis, M. G. Remnant $PbI_2$, an Unforeseen Necessity in High-Efficiency Hybrid Perovskite-Based Solar Cells? *APL Mater.* **2014**, *2* (9), 91101.

(24) Wang, L.; McCleese, C.; Kovalsky, A.; Zhao, Y.; Burda, C. Femtosecond Time-Resolved Transient Absorption Spectroscopy of $CH_3NH_3PbI_3$ Perovskite Films: Evidence for Passivation Effect of $PbI_2$. *J. Am. Chem. Soc.* **2014**, *136* (35), 12205–12208.

(25) Bi, D.; El-Zohry, A. M.; Hagfeldt, A.; Boschloo, G. Unraveling the Effect of PbI2 Concentration on Charge Recombination Kinetics in Perovskite Solar Cells. *ACS Photonics* **2015**, *2* (5), 589–594.

(26) Sheng, C.-X.; Zhai, Y.; Olejnik, E.; Zhang, C.; Sun, D.; Vardeny, Z. V. Laser Action and Photoexcitations Dynamics in PbI2 Films. *Opt. Mater. Express* **2015**, *5* (3), 530.

(27) Flender, O.; Klein, J. R.; Lenzer, T.; Oum, K. Ultrafast Photoinduced Dynamics of the Organolead Trihalide Perovskite $CH_3NH_3PbI_3$ on Mesoporous $TiO_2$ Scaffolds in the 320-920 Nm Range. *Phys. Chem. Chem. Phys.* **2015**, *17* (29), 19238–19246.

(28) Lin, M.-F.; Verkamp, M. A.; Ryland, E. S.; Zhang, K.; Vura-Weis, J. Impact of Spatial Chirp on High Harmonic Extreme Ultraviolet Absorption Spectroscopy of Thin Films. *J Opt Sci Am B* **2016**, *33*, 1986–1992.

(29) Sutherland, J. R.; Christensen, E. L.; Powers, N. D.; Rhynard, S. E.; Painter, J. C.; Peatross, J. High Harmonic Generation in a Semi-Infinite Gas Cell. *Opt. Express* **2004**, *12* (19), 4430–4436.

(30) Corkum, P. B. Plasma Perspective on Strong-Field Multiphoton Ionization. *Phys. Rev. Lett.* **1993**, *71* (13), 1994–1997.

(31) Hayashi, T.; Toyoda, K.; Itoh, M. Absorption Spectra of $PbI_2$ Layered Crystals in 2–100 eV Range. *J. Phys. Soc. Japan* **1988**, *57* (5), 1861–1862.

(32) Brown, F.; Gähwiller, C.; Fujita, H.; Kunz, A.; Scheifley, W.; Carrera, N. Extreme-Ultraviolet Spectra of Ionic Crystals. *Phys. Rev. B* **1970**, *2* (6), 2126–2138.

(33) Robertson, J. Tight Binding Band Structure of $PbI_2$ Using Scaled Parameters. *Solid State Commun.* **1978**, *26* (11), 791–794.

(34) Ahuja, R.; Arwin, H.; da Silva, A. F.; Persson, C.; Osorio-Guillén, J. M.; de Almeida, J. S.; Araujo, C. M.; Veje, E.; Veissid, N.; An, C. Y.; Pepe, I.; Johansson, B. Electronic and Optical Properties of Lead Iodide. *J. Appl. Phys.* **2002**, *92* (12), 7219–7224.

(35) Price, M. B.; Butkus, J.; Jellicoe, T. C.; Sadhanala, A.; Briane, A.; Halpert, J. E.; Broch, K.; Hodgkiss, J. M.; Friend, R. H.; Deschler, F. Hot-Carrier Cooling and Photoinduced Refractive Index Changes in Organic–inorganic Lead Halide Perovskites. *Nat. Commun.* **2015**, *6*, 8420.

(36) Manser, J. S.; Kamat, P. V. Band Filling with Free Charge Carriers in Organometal Halide Perovskites. *Nat. Photonics* **2014**, *8* (9), 737–743.

(37) Santomauro, F. G.; Grilj, J.; Mewes, L.; Nedelcu, G.; Yakunin, S.; Rossi, T.; Capano, G.; Haddad, A. Al; Budarz, J.; Kinschel, D.; Ferreira, D. S.; Rossi, G.; Tovar, M. G.; Grolimund, D.; Samson, V.; Nachtegaal, M.; Smolentsev, G.; Kovalenko, M. V; Chergui, M. Localized Holes and Delocalized Electrons in Photoexcited Inorganic Perovskites:

Watching Each Atomic Actor by Picosecond X-Ray Absorption Spectroscopy. *Struct. Dyn.* **2017**, *4* (4), 44002.

(38) Weathersby, S. P.; Brown, G.; Centurion, M.; Chase, T. F.; Coffee, R.; Corbett, J.; Eichner, J. P.; Frisch, J. C.; Fry, A. R.; Guehr, M.; Hartmann, N.; Hast, C.; Hettel, R.; Jobe, R. K.; Jongewaard, E. N.; Lewandowski, J. R.; Li, R. K.; Lindenberg, A. M.; Makasyuk, I.; May, J. E.; McCormick, D.; Nguyen, M. N.; Reid, A. H.; Shen, X.; Sokolowski-Tinten, K.; Vecchione, T.; Vetter, S. L.; Wu, J.; Yang, J.; D??rr, H. A.; Wang, X. J. Mega-Electron-Volt Ultrafast Electron Diffraction at SLAC National Accelerator Laboratory. *Rev. Sci. Instrum.* **2015**, *86* (7), 1–7.

(39) Singh, N.; Sharma, P. K. Debye-Waller Factors of Cubic Metals. *Phys. Rev. B* **1971**, *3* (4), 1141–1148.

(40) Mannebach, E. M.; Li, R.; Duerloo, K. A.; Nyby, C.; Zalden, P.; Vecchione, T.; Ernst, F.; Reid, A. H.; Chase, T.; Shen, X.; Weathersby, S.; Hast, C.; Hettel, R.; Coffee, R.; Hartmann, N.; Fry, A. R.; Yu, Y.; Cao, L.; Heinz, T. F.; Reed, E. J.; Duerr, H. A.; Wang, X.; Lindenberg, A. M. Dynamic Structural Response and Deformations of Monolayer $MoS_2$ Visualized by Femtosecond Electron Diffraction. *Nano Lett.* **2015**, *15* (10), 6889–6895.

(41) Tubbs, M. R.; Forty, A. J. The Exciton Spectrum of Lead Iodide. *J. Phys. Chem. Solids* **1965**, *26* (4), 711–719.

(42) Blasi, C. De; S., G.; Manfredotti, C.; Micocci, G.; Ruggiero, L.; Tepore, A. Trapping Levels in $PbI_2$. *Solid State Commun.* **1978**, *25*, 149–153.

(43) Levinson, J.; Shepherd, F. R.; Scanlon, P. J.; Westwood, W. D.; Este, G.; Rider, M. Conductivity Behavior in Polycrystalline Semiconductor Thin Film Transistors. *J. Appl. Phys.* **1982**, *53* (2), 1193–1202.

(44) Shockley, W.; Read, W. T. Statistics of the Recombinations of Holes and Electrons. *Phys. Rev.* **1952**, *87* (5), 835–842.

(45) Esser, A.; Seibert, K.; Kurz, H.; Parsons, G. N.; Wang, C.; Davidson, B. N.; Lucovsky, G.; Nemanich, R. J. Ultrafast Recombination and Trapping in Amorphous Silicon. *Phys. Rev. B* **1990**, *41* (5), 2879–2884.

(46) Sun, D.; Rao, Y.; Reider, G. A.; Chen, G.; You, Y.; Brézin, L.; Harutyunyan, A. R.; Heinz, T. F. Observation of Rapid Exciton–Exciton Annihilation in Monolayer Molybdenum Disulfide. *Nano Lett.* **2014**, *14* (10), 5625–5629.

(47) Tanguy, C.; Hulin, D.; Mourchid, A.; Fauchet, P. M.; Wagner, S. Free-Carrier and Temperature Effects in Amorphous Silicon Thin Films. *Appl. Phys. Lett.* **1988**, *53* (10), 880–882.

(48) Lifshitz, E.; Yassen, M.; Bykov, L.; Dag, I.; Chaim, R. Nanometer-Sized Particles of Lead Iodide Embedded in Silica Films. *J. Phys. Chem.* **1994**, *98* (5), 1459–1463.

(49) Makino, T.; Gu, P.; Watanabe, M.; Hayashi, T. Excitation Energy Dependence of Biexciton Formation Efficiency in PbI$_2$. *Solid State Commun.* **1995**, *93* (12), 983–987.

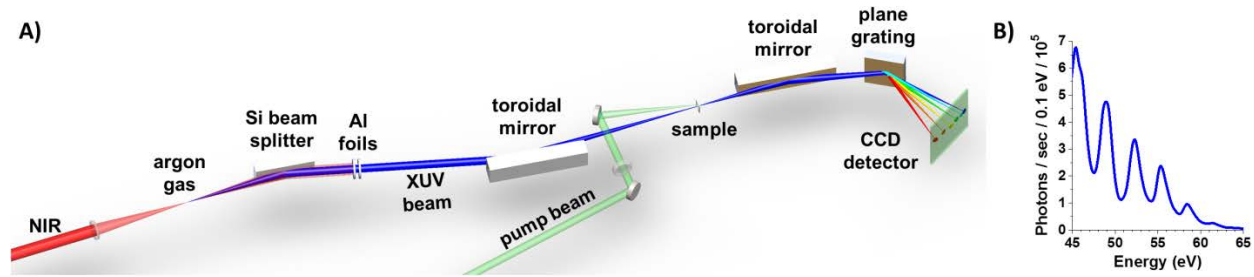

**Figure 1. (A)** Tabletop XUV transient absorption apparatus. An 800 nm laser pulse is focused into an Ar gas cell, where the intense electric field ionizes the Ar atoms, accelerates the free electrons, then drives them back into the Ar ions, releasing XUV photons through the process of high-harmonic generation. A silicon beam splitter followed by Al foils absorbs the residual 800 nm laser pulse. The XUV beam is then focused with a toroidal mirror onto the sample and dispersed onto an array CCD. The 400 nm pump pulse is focused onto the sample at a 5° angle with respect to the XUV pulse. The entire system after the gas cell is maintained at $<10^{-6}$ torr to prevent XUV absorption by air. The sample is raster-scanned and cooled with flowing $N_2$ gas to avoid damage. **(B)** XUV continuum as detected at the CCD after transmission through all XUV optics and the $PbI_2$ sample. Peaks are observed every 3.1 eV, corresponding to the odd harmonics of the 800 nm driving laser pulse, but a broad continuum beneath these peaks allows the 45-62 eV region to be used for absorption spectroscopy.

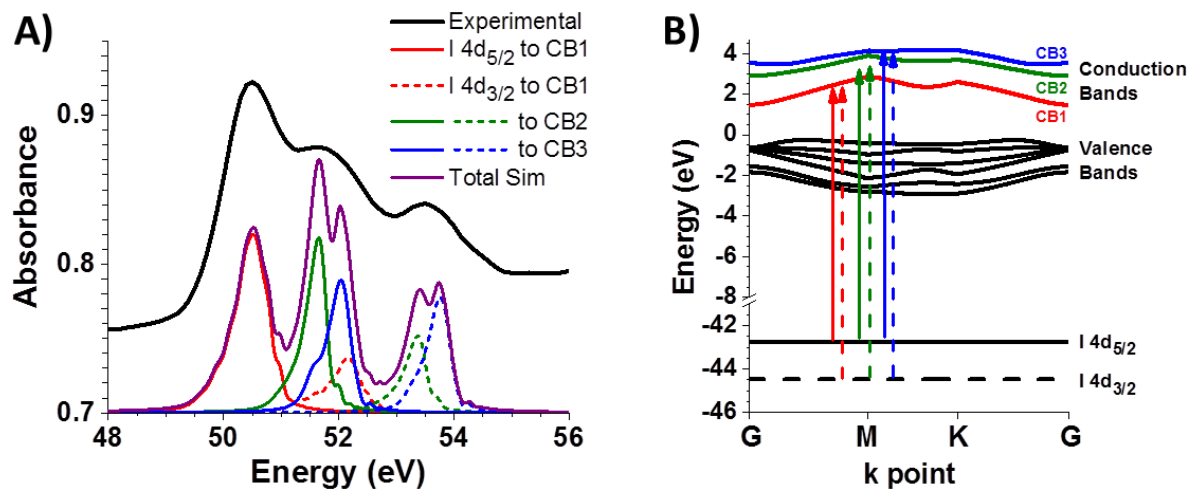

**Figure 2.** (**A**) Ground-state XUV absorption spectrum of PbI$_2$ with DFT simulation. The spectrum is the sum of transitions from the iodine 4$d_{5/2}$ and 4$d_{3/2}$ spin-orbit core levels to the first three empty conduction bands. (**B**) Band structure of PbI$_2$ calculated using DFT, with XUV transitions from the two core levels to the first three conduction bands shown as solid and dashed vertical arrows, respectively. Summation of these transitions over all *k* points produces the simulation shown in (A). The core-level binding energy is underestimated at this level of theory, so the calculated spectrum is manually shifted in energy by +4.7 eV to match the experimental results.

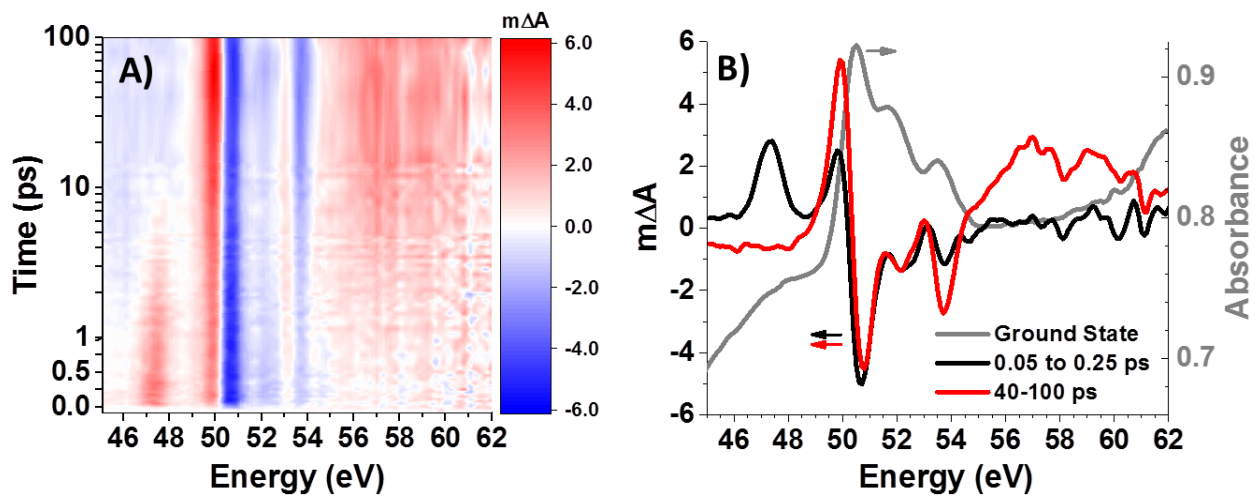

**Figure 3.** (A) XUV transient absorption spectra of PbI$_2$ after photoexcitation at 3.1 eV. (B) Transient absorption spectra of PbI$_2$ at early (averaged from 0.05 to 0.25 ps) and late (40 to 100 ps) delay times, with ground state spectrum for reference. The early-time signal rises within the instrument response and has a positive feature at 47.4 eV and an asymmetric derivative-like feature with positive and negative peaks at 49.9 eV and 50.7 eV, respectively. This spectrum evolves cleanly in ~4 ps to the late-time spectrum, in which the 47.4 eV peak disappears, the peak at 49.9 eV doubles in intensity, and a broad positive feature appears from 55 eV to 62 eV.

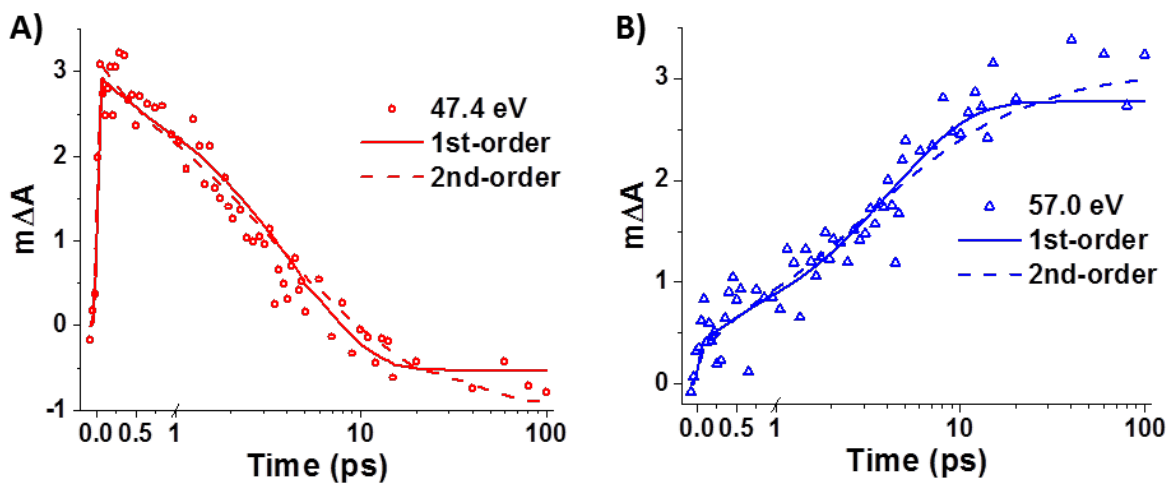

**Figure 4.** Kinetic slices at **(A)** 47.4 eV and **(B)** 57.0 eV, showing the decay of the hole feature and rise of the heat signature (see Section 3.3). Global fits using first- and second-order rate laws are shown as solid and dashed lines, respectively.

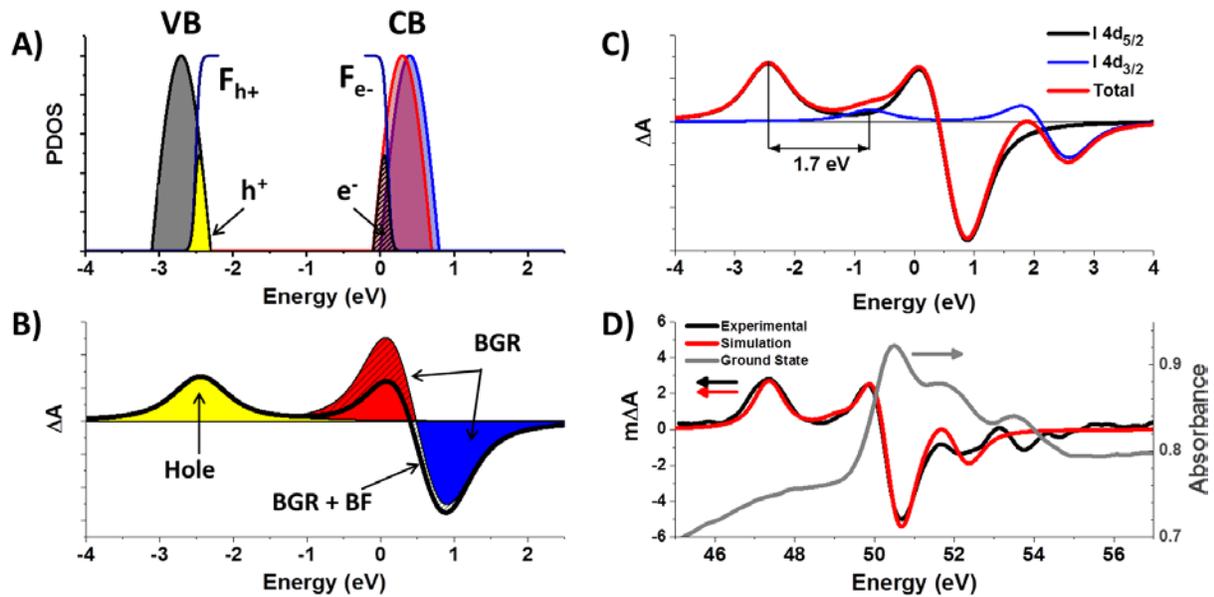

**Figure 5.** Schematic view of the XUV transient absorption spectrum at early times, showing **(A)** the valence and conduction band density of states, with both the bandgap renormalization (BGR) and band-filling (BF) exaggerated for clarity. The conduction band minimum is set as the zero of energy. **(B)** Resulting transient absorption difference spectrum after instrument and lifetime broadening (see text). Photoexcitation produces holes in the valence band with a density of states (yellow) corresponding to the product of the valence band DOS (grey) and a Fermi function $F_{h+}$. This opens up a new core→VB absorption channel and results in a positive transient signal that is redshifted from the CB edge by approximately 2.4 eV. Bandgap renormalization shifts the conduction band to lower energy (blue to red lines in **A**, leading to a derivative-shaped feature in the transient signal (red and blue filled areas in **B**). However, band-filling in the shifted conduction band blocks absorption into the bottom of this band, partially suppressing the induced absorption (hashed areas). **(C)** Inclusion of spin-orbit coupling. The transitions from the I $4d_{3/2}$ core-hole state are shifted by the 1.7 eV spin-orbit splitting and scaled by the relative transition dipole moments calculated with DFT. **(D)** Simulated transient absorption spectrum compared to the experimental short-time transient and the ground-state spectrum.

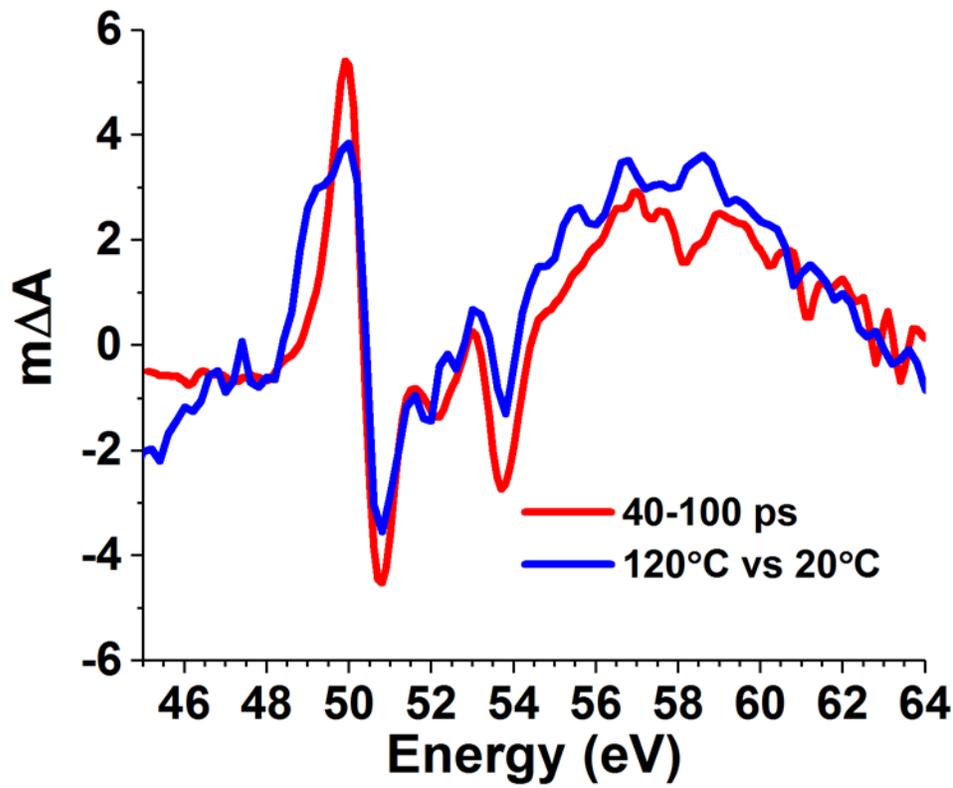

**Figure 6.** Difference spectrum of PbI$_2$ at 120 °C vs 20 °C. The hot spectrum (blue) strongly resembles the late-time transient spectrum (red), suggesting that the initial electronic excitation decays to heat.

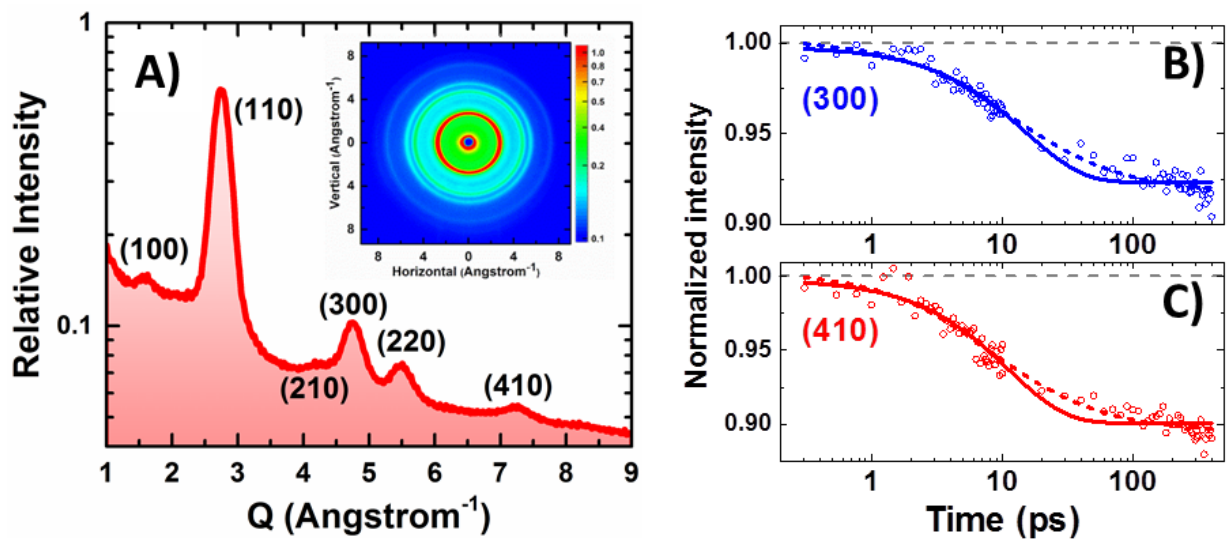

**Figure 7.** **(A)** Radially integrated electron diffraction of 180 nm polycrystalline $PbI_2$ thin film. Due to the transmission geometry of electron diffraction measurement and preferential growth of $PbI_2$ thin film along the C-axis, only in-plane diffraction peaks are observed. Inset: raw diffraction image. **(B)** and **(C)** Debye-Waller thermal response of the (300) and (410) peaks after 3.1 eV photoexcitation. Sold and dashed lines represent first-order and second-order fits, respectively.

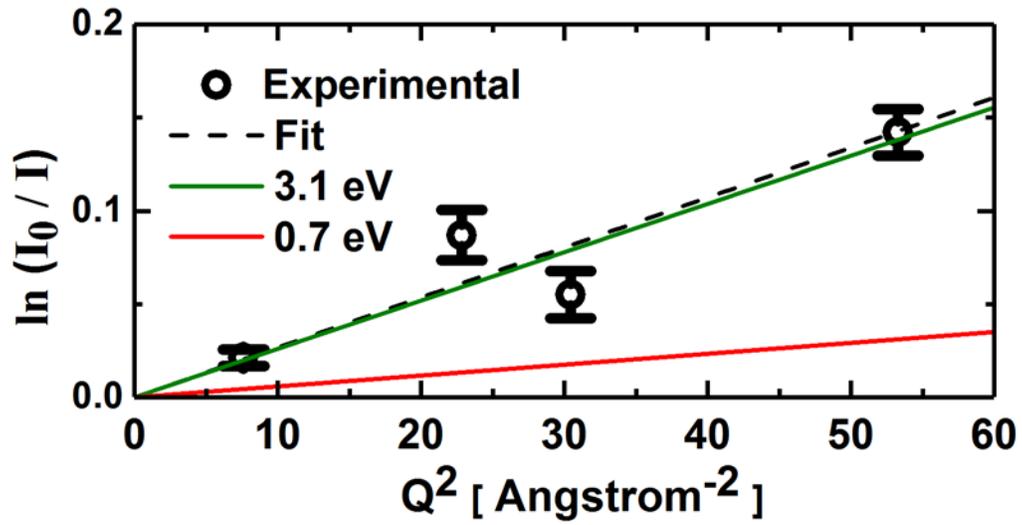

**Figure 8.** Debye-Waller response of 180 nm PbI$_2$ as a function of Q$^2$ after optical excitation at 3.1 eV. The linear fit is shown as the dashed black line. Predicted responses for nonradiative (all 3.1 eV per photon is converted to heat) and nonradiative (only 0.7 eV per photon released as heat) are shown as green and red lines, respectively.

Supplementary Information

Carrier-Specific Femtosecond XUV Transient Absorption of PbI$_2$ Reveals Ultrafast Nonradiative Recombination

Ming-Fu Lin, Max Verkamp, Josh Leveillee, Elizabeth Ryland, Kristin Benke, Kaili Zhang, Clemens Weninger, Xiaozhe Shen, Renkai Li, David Fritz, Uwe Bergmann, Xijie Wang, André Schleife, and Josh Vura-Weis

# Contents

## S1. Sample preparation and characterization

Lead iodide is deposited on silicon nitride ($Si_3N_4$) membranes (Silson, Ltd) using a partially homebuilt thermal evaporation system (Plasmonic Technologies, LLC model LTK350-sys). The substrates are ozone pretreated for ~5 minutes prior to sample coating. For the XUV experiment, 50 nm x 2 mm x 2 mm membranes are used, on 7.5x7.5mm Si frames. This substrate allows us to have 30% XUV transmission in the spectral range from 30 eV to 72 eV. For ultrafast electron diffraction, 100 nm $Si_3N_4$ substrates are used. All the samples are characterized using SEM, XRD and UV-Vis as shown below. Figure S1 shows an image of a thermal evaporated $PbI_2$ thin film (~100 nm) measured by Hitachi S4700 High Resolution SEM. Small platelets of $PbI_2$ crystals are observed in this image.

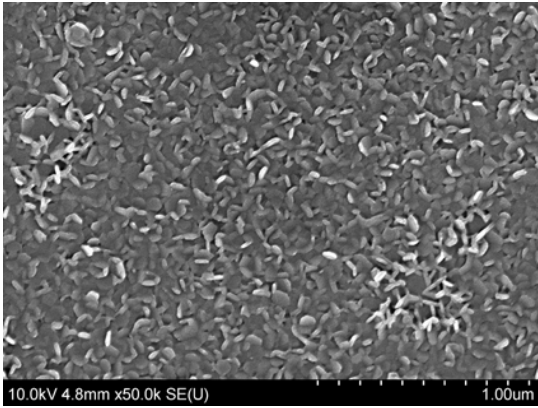

**Figure S1:** SEM of $PbI_2$ film

Figure S2 displays the $PbI_2$ film x-ray diffraction. This is measured by Panalytical/Philips X'pert XRD machine at Cu K-α emission line at 1.5418 Å with data acquisition time of 20 minutes. The preferential growth along $c$-axis is observed. The obtained lattice constant $c$ is ~6.99 Å, consistent with literature values.

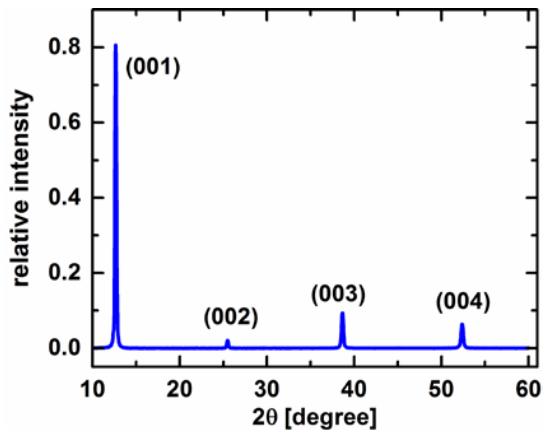

**Figure S2:** X-ray diffraction of $PbI_2$ film.

Figure S3 shows the UV-VIS absorption spectrum of a ~100 nm thick $PbI_2$ on a glass substrate measured by Cary-500-Scan spectrometer. A clear absorption edge located ~500 nm can be observed. A small increase past 510 nm results from interference effects of the thin film.

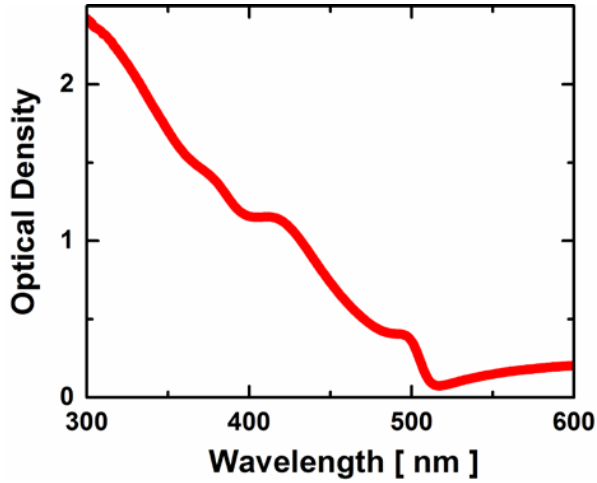

**Figure S3:** UV-Vis of $PbI_2$ film

Figure S4 illustrates an AFM (Asylum Research MFP-3D) measurement of a representative $PbI_2$ film on $Si_3N_4$ membrane. The thickness is 120±10 nm. With this information and the measured reflectivity of $PbI_2$, we are able to deduce an absorption coefficient of ~$0.88 \times 10^5$ cm$^{-1}$ at 3.1 eV, which is consistent with the literature value in the spectral range from 350 nm to 450 nm.[1]

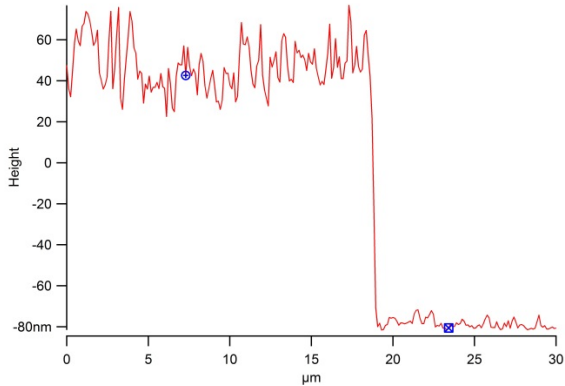

**Figure S4:** AFM profile of $PbI_2$ film.

## S2. PbI$_2$ XPS spectrum and calculation of core-hole lifetime broadening

X-ray photoelectron spectroscopy was used to estimate the core-hole lifetime broadening, under the assumption that the lifetime of the 4d$^{-1}$ state is only minimally perturbed by the conduction-band population (i.e. the lifetime of the core-ionized state after XPS is close to the lifetime of the core-to-valence excited state after XUV absorption).

Figure S5 shows the x-ray photoelectron spectrum of the PbI$_2$ film. XPS was performed on a Kratos Axis ULTRA instrument with a monochromatic Al Kα X-ray source (source energy 1486.61 eV) and a hybrid spherical capacitor energy analyzer. Spectra were collected with an analyzer pass energy of 10 eV. Binding energy scales of PbI$_2$ spectra were calibrated by referencing the C 1s peak from adventitious carbon to 284.8 eV. Reference spectra of a piece of gold-coated silicon wafer were also collected on the same instrument.

The instrumental (Gaussian) resolution of the XPS experiment was estimated to be 0.45 eV by fitting the Au 4f XPS features to Voigt functions with the Lorentzian width (FWHM) constrained to be 0.25 eV.[2] The I 4d photoelectron features (Tougaard baseline subtracted) were then fitted to Voigt functions with the Gaussian σ's constrained to be 0.45 eV. Intrinsic core-hole line width was estimated to be 0.48 eV by taking the average of the Lorentzian widths of the I 4d$_{5/2}$ and 4d$_{3/2}$ peaks.

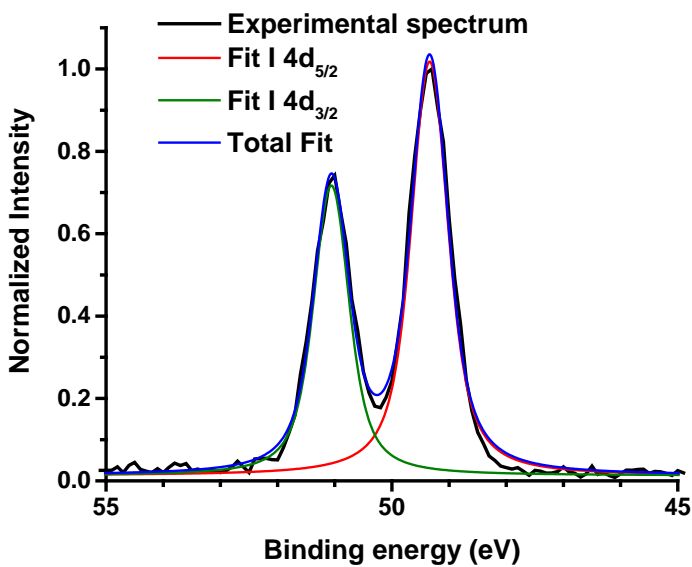

**Figure S5:** X-ray photoelectron spectrum of PbI$_2$ film with Voight fit.

## S3. Gas cooling in XUV transient absorption

A stream of $N_2$ is passed over the sample during the XUV transient absorption experiment to carry away heat from the pump beam. With this gas flow, the XUV and UV-VIS absorption spectra are both unchanged at pump fluence below the damage threshold of ~4.5 mJ/cm$^2$. Spectra taken before and after transient absorption are unchanged.

Without this gas cooling, there is a pronounced pre-time-zero signal due to pump heating that lasts beyond the 1 ms repetition period of the laser. Figure S6 shows this effect. The pre-t0 spectral feature (red) disappears with $N_2$ gas flows through the thin film surface (blue).

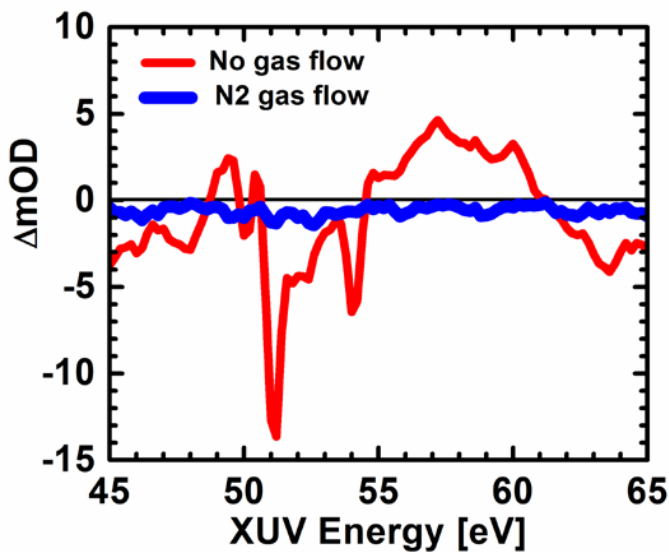

**Figure S6:** Pre-t0 signal with gas flow (blue) and without (red).

## S4. XUV transient absorption with band-edge excitation

The early-time spectrum after pumping at 400 nm (3.1 eV) has upward and downward peaks separated by 3.1 eV. The main text shows that this spectrum can be interpreted as the combined effects of bandgap renormalization and band filling. However, we also considered an alternative interpretation, that of an electron and a hole distribution separated by the 3.1 eV pump energy (with no bandgap renormalization). If this were true, then photoexcitation at 500 nm (2.5 eV) would create hole (upward) and electron (downward) transient peaks separated by 2.5 eV. We tested this by photoexciting the $PbI_2$ sample with the 500 nm, 50 fs output of a noncollinear optical parametric amplifier (TOPAS-White). Figure S7 shows the resulting transient spectrum, averaged between 100 and 200 fs. The spectrum with 500 nm pumping is identical within the noise to that with 400 nm pumping. In particular, the peak positions do not change depending on the pump energy (at least at these delay times), disproving this band-filling only interpretation and supporting the "bandgap renormalization plus band-filling" description presented in the main text. The 400 nm experiment was pumped with a peak fluence of 1.1 mJ/cm$^2$, giving a carrier density of $0.6\times10^{20}$. The 505 nm experiment was pumped at 7.7 mJ/cm$^2$ to account for the lower absorbance at the band edge, giving a carrier density of $0.8\times10^{20}$. At these similar carrier densities, the two transient spectra are equivalent within the noise. Note that the carrier density is calculated here as ((# photons absorbed within FWHM)/(excitation volume)).

The fact that the 3.1 eV pump creates upward and downward peaks separated by 3.1 eV is simply coincidental: 1) The upward hole feature appears 2.4 eV (the band gap) below the CB edge. 2) The first conduction band is ~0.7 eV wide, so when the entire band shifts to lower energy the negative transient appears 0.7 eV above the CB edge. 3) The -2.4 eV position of the upward signal and +0.7 eV position of the downward signal combine to create a peak-to-peak spacing of 3.1 eV.

We also note that the "electron and hole separated by 3.1 eV" concept is unphysical, as the carriers will thermalize to form Fermi distributions at the band edge within the 95 fs instrument response.[7]

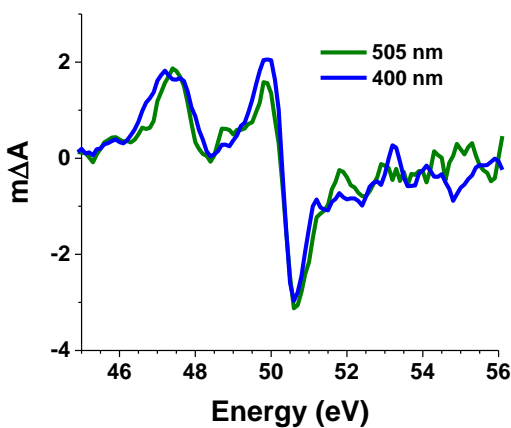

**Figure S7:** XUV transient absorption with 500 nm vs 400 nm pumping.

## S5. Further details on XUV transient absorption experiment

The XUV transient absorption experiment was conducted over four calendar days, with a total of 17 hours of data collection (plus alignment, etc), and the spectra shown are the averaged data from this period (approx 30 minutes per timepoint). Pump/probe spectra were collected with 1 second integration times, and the sample was moved to a new spot after each (pump off/pump on) cycle. No sample damage was observed from either the pump or probe beams. The pump spot size was 200 μm FWHM, and pump/probe overlap was verified with vertical and horizontal knife-edge scans after every 1-2 hours of data collection.

As noted in the main text, singular value decomposition and global fitting to a sequential A→B model was performed using both first and second-order kinetics ($-\frac{d[A]}{dt} = \frac{d[B]}{dt} = k_n[A]^n$), convolved with a Gaussian instrument response function.

Figure S8 compares the averaged initial (0.05 to 0.25 ps) and final (40 to 100 ps) spectra to the species-associated spectra obtained by global fitting to first-order or second-order kinetic models. The SAS of the two components are nearly identical to each other and to the initial and final averages.

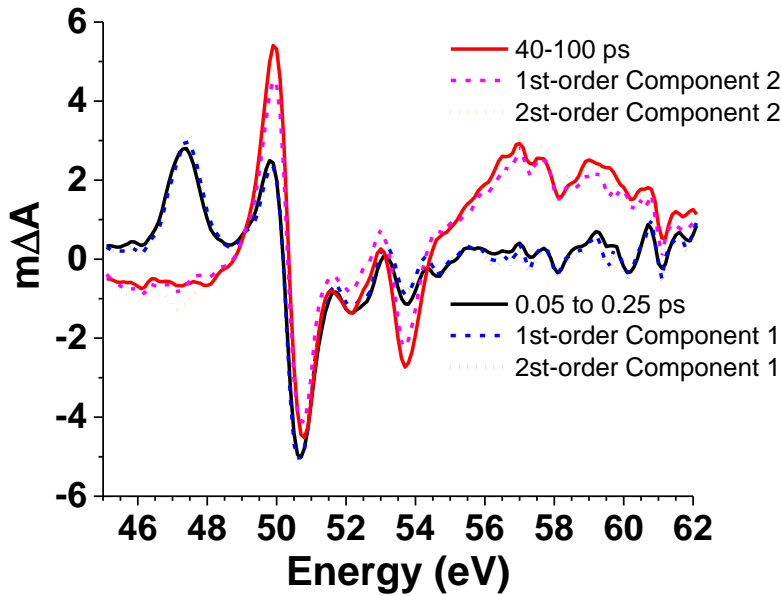

**Figure S8:** Early- and late-time spectra compared to species-associated spectra obtained using first-order or second-order kinetic models.

The main text reports both first-order and second-order rate constants extracted from the respective fits. The second-order differential equation integrates to $N_{(t)} = \frac{N_0}{1+k_2 \times N_0 \times t}$, in which $N_{(t)}$, $N_0$ and $k_2$ are the time-dependent carrier density, initial carrier density and second-order recombination rate constant, respectively. The $k_2$ rates reported in the main text therefore depend on the initial carrier concentration, which was estimated from the measured pump power, reflectivity, and transmission of the sample at 400 nm (i.e. the fitted parameter is "$k_2 \times N_0$", and the estimated $N_0$ is used to recover $k_2$). Note that the actual fits include a convolution with a Gaussian instrument response, which is not shown in the $N_{(t)}$ equation above.

Figure S9 shows the experimental transient absorption contour plot, along with the reconstructed contours from the first-order and second-order models. The residual contour plots are also shown.

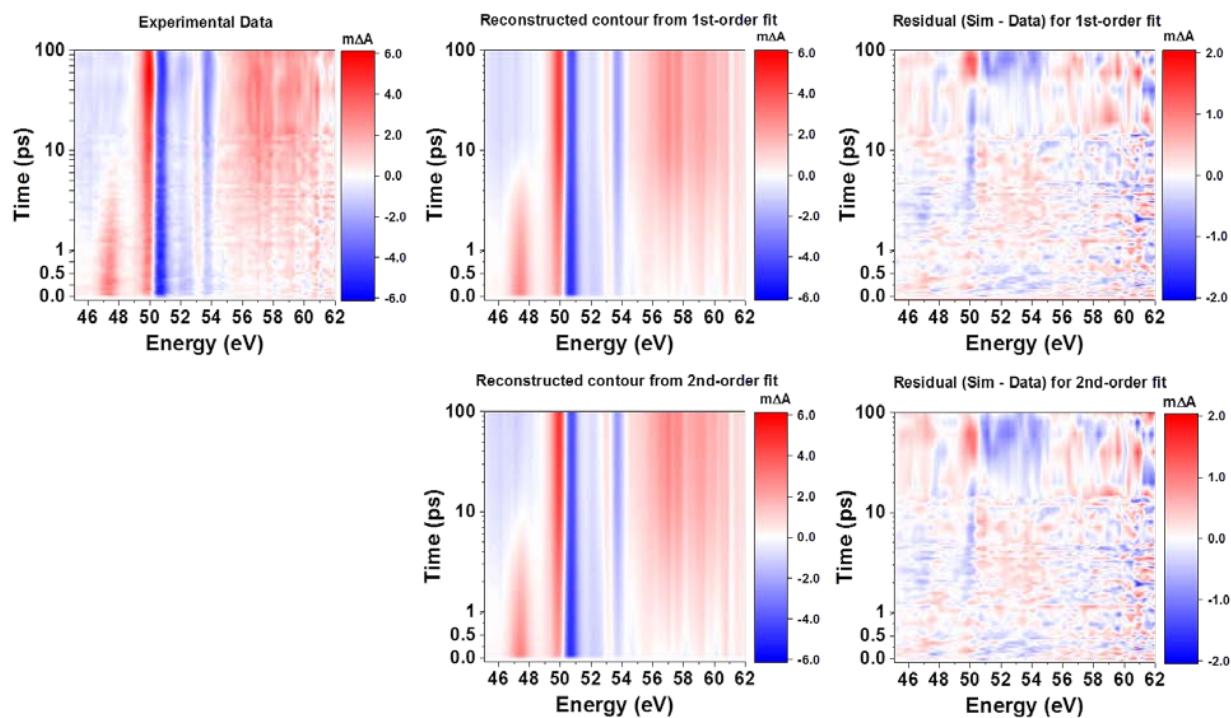

**Figure S9:** Experimental XUV transient absorption contour plot, reconstructed contour plots from 1st and 2nd-order fits, and residuals from each fit. Note that the Z scale of the residuals is 1/3 that of the data and reconstructions.

Both the first-order and second-order fit residuals show a small non-random mismatch between the fit and the experiment between 49 and 55 eV. This is shown more clearly in Figure S10, which shows the early-time and late-time averaged spectra along with slices of the 2nd-order fit residual at 0.5 ps, 3.0 ps, and 40 ps. The error in the fit at 3 ps and 40 ps is approximately 1 mΔA, and is likely caused by a time-dependent shift in the conduction band edge as carrier recombination heats the lattice. More detailed deconvolution of this shift was not possible given the noise level of the data.

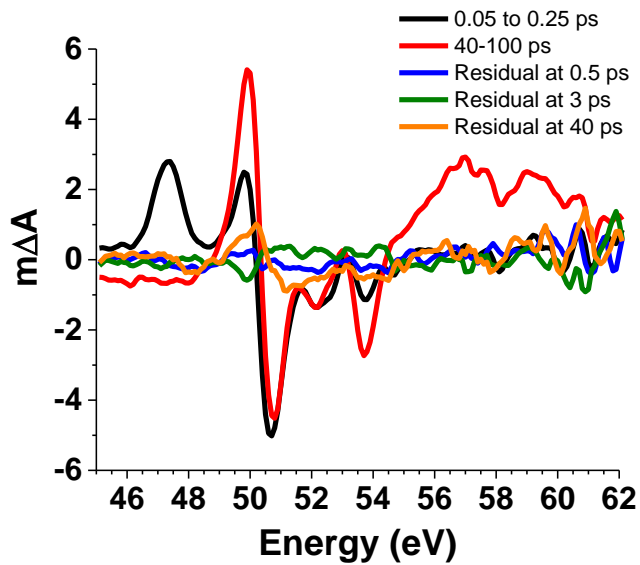

**Figure S10**: Experimental XUV transient absorption contour plot, reconstructed contour plots

There is one additional detail that is not captured in the fits: a slight time-dependent shift in the positive feature at ~47.4 eV. This shift is analyzed by fitting the 46.0 eV to 48.5 eV region with a Gaussian function, as shown in Figure S11A. In the first 100 fs, as the pump pulse builds the excited-state population, this feature redshifts from 47.5 eV to 47.3 eV. Over the next few picoseconds, the peak then blueshifts to 47.6 eV.

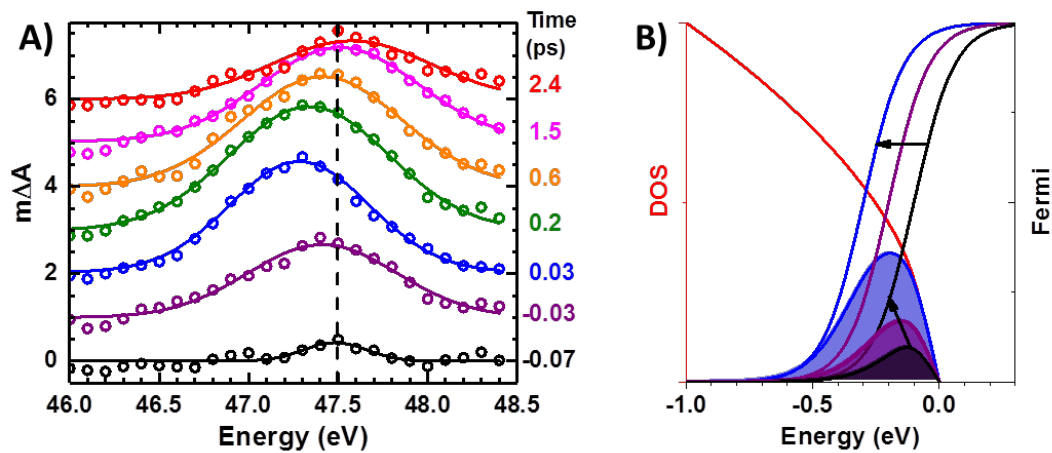

**Figure S11**: **(A)** Gaussian fits of the transient spectral feature near 47.4 eV (corresponding to the photoinduced hole population) within the first 2.4 ps delay time range. The vertical dashed line at 47.5 eV marks the peak center at rising edge of the pump pulse at -70 fs. The peak energy redshifts to 47.3 eV during the 95 fs instrument response, then blueshifts to 47.6 eV as the holes recombine. **(B)** Model conduction band density of states and Fermi functions, with the zero of energy set here to the valence band maximum. As the pump pulse creates holes in the valence band, the peak of the hole distribution (DOS * Fermi) shifts to lower energy.

This shift is a straightforward consequence of the model discussed in Section 3.2 and Figure 5 of the main paper. As the pump pulse creates holes in the valence band, the quasi-Fermi level shifts to lower energy. The peak of the hole distribution therefore redshifts, which is reflected in the position of the XUV absorption. This is shown pictorially in Figure S11B. Subsequently, the hole population decays upon carrier recombination and the quasi-Fermi level shifts back to higher energy, causing the corresponding blueshift in the XUV signal. As discussed in Section 3.3 of the main text, sample heating upon carrier recombination shrinks the band gap, leading to a small net blueshift of this peak beyond 2 ps. Note that the XUV absorption signal includes instrument (Gaussian) and lifetime (Lorentzian) broadening, leading to an effectively Gaussian lineshape in Figure S11A despite the asymmetry of the hole distributions shown in Figure S11B.

## S6. Transient absorption modeling

Bandgap renormalization was estimated by extrapolating from values measured[3] in $CH_3NH_3PbI_3$, which has a similar dielectric constant $(25.7)^4$ as $PbI_2$ $(20.8)^5$. The band gap reduction $\Delta E_g$ is proportional to the excitation density $(n_0^{1/3})$ and inversely proportional to the static dielectric constant.[6] The range of excitation density for our experiment is $n_0 = 10^{19}$ to $10^{20}$ cm$^{-3}$ which corresponds to $\Delta E_g$ values from 40 to 100 meV. For our simple model, the shift is assumed to be the same for both valence and conduction bands, giving a change in position of each band of 20 to 50 meV. The final shift value used in our model for each band (30 meV) is within this range. As noted in the text, there is likely a shift in the I 4d core level after photoexcitation, which will counteract the effect of the band shift and reduce the apparent magnitude of bandgap renormalization.

Carriers (electrons and holes) were assumed to have reached a thermal distribution from electron-electron and hole-hole interactions within our instrument response (95 fs).[7] Thus, a Fermi distribution with carrier temperature $T_c$ and quasi-Fermi levels of $E^q_{fe}$ and $E^q_{fh}$ was used to represent the band-filling (hole burning). With our experimental broadening (Voigt with Gaussian FWHM of 0.4 eV and Lorentzian FWHM of 0.5 eV), the impact of $T_c$ will be difficult to observe so modeling was done using 300K. The amount of filling was controlled by shifting the quasi-Fermi levels away from the valence band maximum and conduction band minimum to match the magnitude of the valence band feature. The fill factor (amount filled over full area of conduction/valence band) used for the best fit was 1.7%. The DFT band structure calculation gives a total partial density of states in the first conduction band of $1.34 \times 10^{22}$ cm$^{-3}$. Given a photoinduced carrier density of $1.5 \times 10^{20}$ cm$^{-3}$, the experimental band filling is 1.1%, an excellent match for the fit result.

Figure S12 displays the model conduction band used for the simulations before and after broadening as discussed above. The bottom of the unbroadened band is set to 49.8 eV so that the rising edge of the broadened band aligns with that of the experimental spectrum.

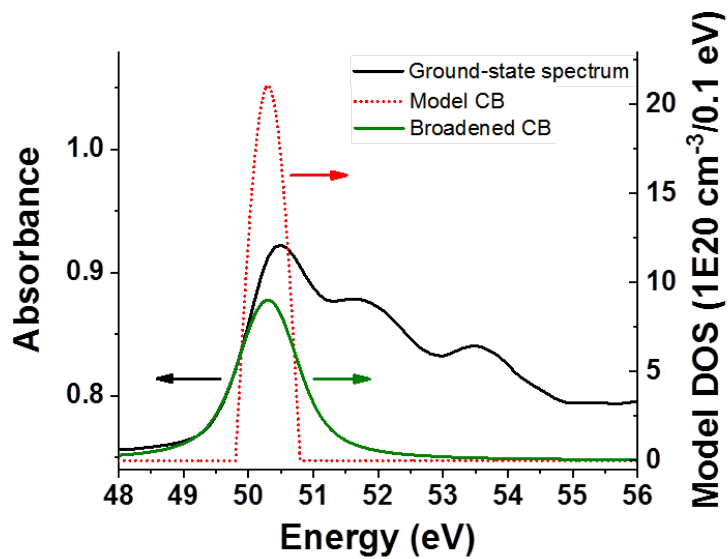

**Figure S12**: Model conduction band before and after broadening, along with the ground-state absorption spectrum.

Figure S13 and Figure S14 show the simulated transient spectra when considering only bandgap renormalization (BGR) or only band-filling (BF). Spin-orbit splitting is not included for simplicity. Neither model resembles the experimental spectrum, showing that both BGR and BF are required to explain the observed spectrum.

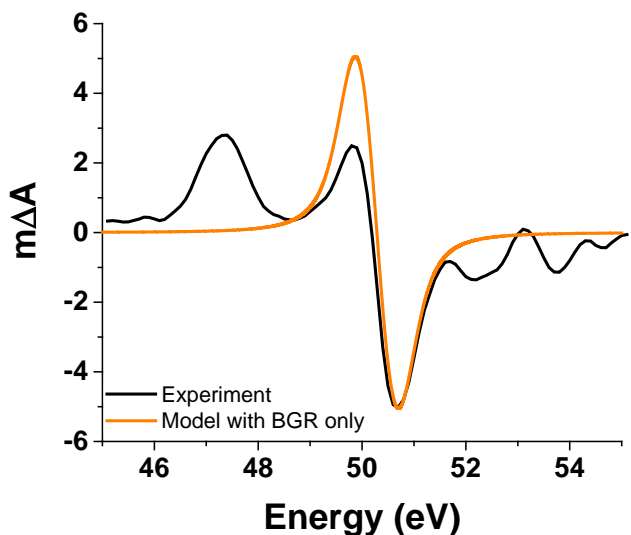

**Figure S13**: Experimental transient spectrum at early times (0.05 ps to 0.25 ps) compared to model with only bandgap renormalization, using a band shift of 30 meV

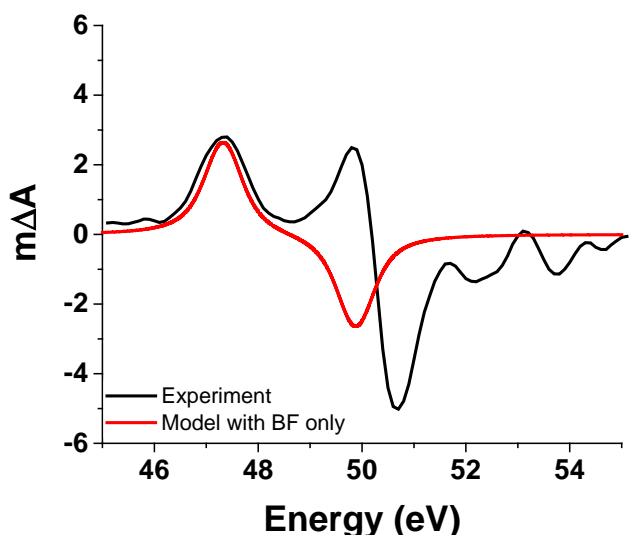

**Figure S14**: Experimental transient spectrum at early times compared to model with only band-filling, using 1.7% filling of the conduction band.

### S7. Ultrafast Electron Diffraction (UED) measurement

We utilized ultrafast electron diffraction to further characterize temperature jumps of $PbI_2$ from optical laser excitation at 400 nm. Due to the normal incidence of UED in transmission mode, only (h,k,0) in-plane diffractions are measured and this result agrees with the observation of XRD measurements which shows a preferential growth of thin film along the c-axis (see Figure S2). An electron beam density of ~10 fC per pulse is employed for the experiments. The experiment is performed at a 180 Hz repetition rate and with 2 second exposure time per delay time point for the detector. Each diffraction pattern contains several rings, corresponding to different momentums Q ($Q = 2\pi/d$). By taking the diffraction images as a function of pump-probe delay time, we follow the individual Q diffraction intensity change over a delay time window. The suppression of diffraction intensity as a function of delay time provides us the Debye-Waller response that measures the mean-square displacement of atoms in the thin film.[8]

## S8. Analysis of Debye-Waller response

The Debye-Waller response is obtained from fits of $\ln(I_0/I)$ versus $Q^2$ and used to estimate the sample temperature jump $\Delta T$ after photoexcitation. $I_0$ and $I$ are the diffraction intensities at far negative and +100 ps delay times. This experimental $\Delta T$ is compared to the temperature jump calculated from the known pump fluence at 3.1 eV (400 nm), absorption cross section, reflectivity between interfaces, film thickness and specific heat. Figure 9 of the main text shows the experimental response of a 180 nm $PbI_2$ film after irradiation with a carrier density of $0.33\times10^{20}$ cm$^{-3}$, along with predicted responses if (1) the full 3.1 eV photon energy is converted to heat (nonradiative recombination) or (2) The carriers relax to the band edge and release 0.7 eV of energy as heat, then fluoresce (radiative recombination). Figure S15 reproduces this figure and adds an additional experiment, in which the $PbI_2$ film was irradiated with a carrier density of $1.2\times10^{20}$ cm$^{-3}$. In both cases, the experimental Debye-Waller response is an excellent match for the nonradiative prediction, with a slope that is much larger than is predicted for a radiative mechanism. See below for a detailed description of the pump fluence and carrier density.

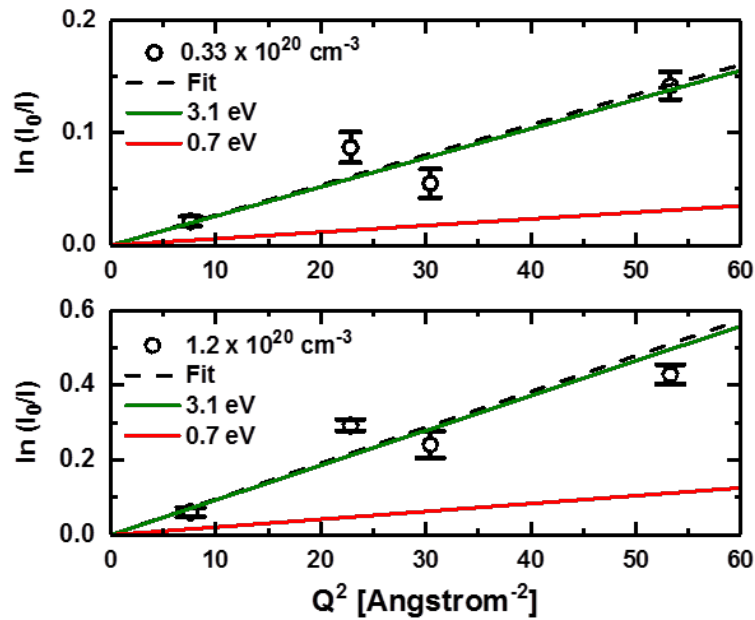

**Figure S15:** Debye-Waller responses from UED measurements at excitation densities of $0.33\times10^{20}$ cm$^{-3}$ and $1.2\times10^{20}$ cm$^{-3}$. Experimental Debye-Waller response is shown as circles. Linear fit is shown as blacked dashed line, with slope given in Table S1. The predicted response assuming that 3.1 eV (nonradiative) or 0.7 eV (radiative) of energy per absorbed photon is converted to heat is shown as green and red lines, respectively.

The Debye-Waller response measures the difference in mean-square displacement between room temperature and final temperature. The final temperature is a sum of the temperature jump and room temperature.

$\ln \frac{I_0}{I} = \ln \frac{I_{300K}}{I_T} = \Delta \langle u^2 \rangle \times Q^2$, where slope $\Delta \langle u^2 \rangle = \langle u^2 \rangle_T - \langle u^2 \rangle_{300K}$

The mean square displacement at temperature T can be calculated[9] as $\langle u^2 \rangle_T = \frac{3\hbar^2}{mk_B\theta_D}\left(\frac{\Phi_{(Z)}}{Z} + \frac{1}{4}\right)$, where $\hbar = \frac{h}{2\pi}$, $k_B$ the Boltzmann constant, $m$ the averaged mass and $\theta_D$ the Debye temperature. $\Phi_{(Z)}$ can be evaluated using analytical function shown as follows.

$\Phi_{(Z)} = \frac{1.6449}{Z}(1 - e^{-0.64486 \times Z})$ with $z = \frac{\theta_D}{T}$.[10]

Here, we use $\theta_D = 99.4\ K$ and averaged mass $m = \frac{201+127}{2} = 164$, respectively.[11] Table S1 summarizes the change in temperature $\Delta T$ calculated from the fitted slopes in Figure S7, as well as the predicted $\Delta T$ for nonradiative and radiative recombination (see below)

Table S1: Temperature jump calculated from input pump energy and absorbance

| Incident energy (μJ) | Absorbed energy (μJ) | Carrier Density (cm$^{-3}$) | Fitted Slope (Å$^2$) | Fitted $\Delta T$ (K) | Nonradiative $\Delta T$ (K) | Radiative $\Delta T$ (K) |
|---|---|---|---|---|---|---|
| 1.35 | 0.89 | 0.33×10$^{20}$ | 0.00268 | 28 ± 3 | 27.5 | 6.2 |
| 4.8 | 3.17 | 1.2×10$^{20}$ | 0.00957 | 101 ± 10 | 98.0 | 22.1 |

The carrier density and predicted temperature jump is calculated from the measured absorbance, incident pump energy and estimated reflectance. The complex refractive index of PbI$_2$ at 400 nm is 3.22-1.32i. With a stacked thin film model[12] that considers interface reflection between vacuum/PbI$_2$ and PbI$_2$/Si$_3$N$_4$, we obtain absorption of ~66%. Note that in the UED experiment, the pump pulse was incident on the PbI$_2$ side of the sample, while in the XUV experiment the pump pulse was incident on the Si$_3$N$_4$ side, resulting in only 6% reflective losses in the latter.

The specific heat and density of PbI$_2$ are 0.168 J/g/K and 6.16 g/cm$^3$, respectively. The 100 nm thick Si$_3$N$_4$ substrate has a specific heat of 0.17 J/g/K and density of 3.44 g/cm$^3$, and it is assumed that thermal equilibrium between the PbI$_2$ and Si$_3$N$_4$ is complete by 100 ps after photoexcitation (when the final Debye-Waller response is measured).

The predicted $\Delta T$, assuming nonradiative recombination, is calculated using:

$$absorbed\ energy = \Delta H = m \times s \times \Delta T$$

where m is the mass of the PbI$_2$ and Si$_3$N$_4$ layers (calculated from the densities above, the 410 nm pump spot size, and the thickness of each layer) and s is the specific heat of each layer. Calculated values of $\Delta T$ assuming the that the full absorbed energy (3.1 eV per photon) is converted to heat are shown in Table S1. $\Delta T$ is also calculated in the case that the electron and hole relax 0.7 eV in total to the band edge (with that energy released as heat), then recombine radiatively. The excellent agreement between between the experimentally fitted $\Delta T$ and the nonradiative prediction indicates that the recombination is primarily nonradiative.

## S9. Band structure calculations

Density functional theory (DFT) calculations were performed using the VASP code.[13,14] A Perdew-Burke-Ernzerhof (PBE)[15] generalized gradient approximation (GGA) was used to treat exchange and correlation. Electron-ion interaction is described using the projector-augmented wave method.[16,17] Iodine 4d states are treated as valence electrons. Cell volume and atomic positions are relaxed until the Hellmann-Feynman forces are below 5E-3 eV/A. Cell energies are converged using a plane-wave cutoff energy of 750 eV and a 7x7x7 Monkhorst-Pack k-point grid. Spin-orbit coupling is included in the.[18] Optical transition matrix elements and the dielectric function are calculated on a denser 13x13x13 gamma-centered k-point grid using the PAW framework and the longitudinal.[19]

Within DFT, the electronic gap of $PbI_2$ in the 2H polymorph is 2.46 eV, and is indirect between the H (valence band maximum) and A (conduction band minimum) point in the BZ (<0.0, 0.0, 0.5> and <0.33, 0.33, 0.5> in Cartesian coordinates respectively). This band gap is reduced to 1.71 eV due to the strong spin-orbit effect induced by Pb and I. The iodine 4d states, all located at -43.5 eV below the valence band edge, split by ~1.7 eV between the $4d_{3/2}$ states (at -44.4 eV) and the $4d_{5/2}$ states (at -42.7 eV). Figure S16 shows the calculated band structure of $PbI_2$ (only a subset is shown in Figure 1B of the main text)

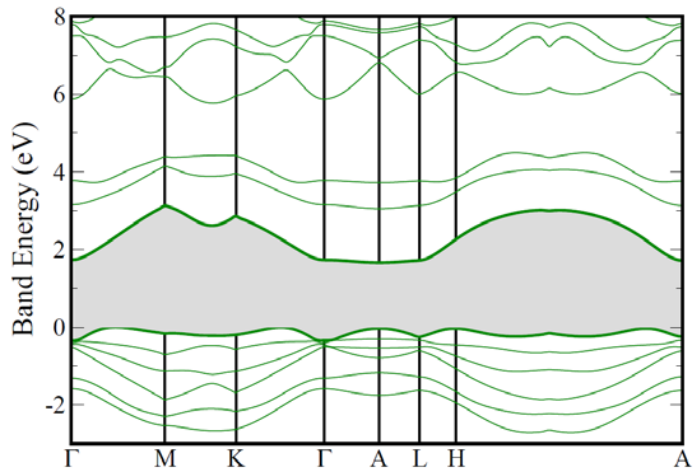

**Figure S16:** Full band structure of $PbI_2$

REFERENCES
(1) Blasi, C. De; S., G.; Manfredotti, C.; Micocci, G.; Ruggiero, L.; Tepore, A. Trapping Levels in $PbI_2$. *Solid State Commun.* **1978**, *25*, 149–153.
(2) Lindau, I.; Pianetta, P.; Yu, K.; Spicer, W. E. The Intrinsic Linewidth of the 4f Levels in Gold as Determined by Photoemission. *Phys. Lett. A* **1975**, *54* (1), 47–48.
(3) Yang, Y.; Ostrowski, D. P.; France, R. M.; Zhu, K.; van de Lagemaat, J.; Luther, J. M.; Beard, M. C. Observation of a Hot-Phonon Bottleneck in Lead-Iodide Perovskites. *Nat. Photonics* **2015**, *10* (1), 53–59.
(4) Frost, J. M.; Butler, K. T.; Brivio, F.; Hendon, C. H.; van Schilfgaarde, M.; Walsh, A. Atomistic Origins of High-Performance in Hybrid Halide Perovskite Solar Cells. *Nano Lett.* **2014**, *14* (5), 2584–2590.
(5) Young, K. F.; Frederikse, H. P. R. Compilation of the Static Dielectric Constant of Inorganic Solids. *J. Phys. Chem. Ref. Data* **1973**, *2* (2), 313–410.
(6) Bennett, B. R.; Soref, R. A.; Del Alamo, J. A. Carrier-Induced Change in Refractive Index of InP, GaAs and InGaAsP. *IEEE J. Quantum Electron.* **1990**, *26* (1), 113–122.
(7) Richter, J. M.; Branchi, F.; Valduga de Almeida Camargo, F.; Zhao, B.; Friend, R. H.; Cerullo, G.; Deschler, F. Ultrafast Carrier Thermalization in Lead Iodide Perovskite Probed with Two-Dimensional Electronic Spectroscopy. *Nat. Commun.* **2017**, *8* (1), 376.
(8) Mannebach, E. M.; Li, R.; Duerloo, K. A.; Nyby, C.; Zalden, P.; Vecchione, T.; Ernst, F.; Reid, A. H.; Chase, T.; Shen, X.; Weathersby, S.; Hast, C.; Hettel, R.; Coffee, R.; Hartmann, N.; Fry, A. R.; Yu, Y.; Cao, L.; Heinz, T. F.; Reed, E. J.; Duerr, H. A.; Wang, X.; Lindenberg, A. M. Dynamic Structural Response and Deformations of Monolayer $MoS_2$ Visualized by Femtosecond Electron Diffraction. *Nano Lett.* **2015**, *15* (10), 6889–6895.
(9) Harb, M.; Enquist, H.; Jurgilaitis, A.; Tuyakova, F. T.; Obraztsov, A. N.; Larsson, J. Phonon-Phonon Interactions in Photoexcited Graphite Studied by Ultrafast Electron Diffraction. *Phys. Rev. B - Condens. Matter Mater. Phys.* **2016**, *93* (10), 1–7.
(10) Hardy, K. A.; Parker, F. T.; Walker, J. C. A Better Approximation to the Debye-Waller Factor. *Nucl. Instruments methods* **1970**, *86* (7), 171–172.
(11) Sears, W. M.; Morrison, J. A. Low Temperature Thermal Properties of $PbI_2$. *J. Phys. Chem. Solids* **1979**, *40* (7), 503–508.
(12) Heavens, O. S. Optical Properties of Thin Solid Films. In *Optical Properties of Thin Solid Films*; Dover Publications, INC, 1965.
(13) Kresse, G.; Furthmüller, J. Efficiency of Ab-Initio Total Energy Calculations for Metals and Semiconductors Using a Plane-Wave Basis Set. *Comp. Mater. Sci.* **1996**, *6*, 15–50.
(14) Kresse, G.; Furthmüller, J. Efficient Iterative Schemes for Ab Initio Total-Energy Calculations Using a Plane-Wave Basis Set. *Phys. Rev. B* **1996**, *54*, 11169–11186.
(15) Perdew, J. P.; Burke, K.; Ernzerhof, M. Generalized Gradient Approximation Made Simple. *Phys. Rev. Lett.* **1996**, *77*, 3865–3868.
(16) Blöchl, P. E. Projector Augmented-Wave Method. *Phys. Rev. B* **1994**, *50*, 17953–17979.
(17) Kresse, G.; Joubert, D. From Ultrasoft Pseudopotentials to the Projector Augmented-Wave Method. *Phys. Rev. B* **1999**, *59*, 1758–1775.
(18) Hobbs, D.; Kresse, G.; Hafner, J. Fully Unconstrained Noncollinear Magnetism within the Projector Augmented-Wave Method. *Phys. Rev. B* **2000**, *62*, 11556–11570.
(19) Gajdoš, M.; Hummer, K.; Kresse, G.; Furthmüller, J.; Bechstedt, F. Linear Optical Properties in the Projector-Augmented Wave Methodology. *Phys. Rev. B* **2006**, *73*, 45112.